\documentclass[twocolappendix]{emulateapj}
\usepackage{subfigure}
\usepackage{amsmath}

\newcommand{\bmath}[1]{\mbox{\boldmath{$#1$}}}

\newcommand{\del}{{\bf \nabla}}
\newcommand{\alf}{{\rm Alfv\acute{e}n}}
\newcommand{\cs}{c_{\rm s}}
\newcommand{\va}{v_{\rm A}}

\begin{document}

\title{Turbulence in the Outer Regions of Protoplanetary Disks. \\ I. Weak Accretion with No Vertical Magnetic Flux}

\author{Jacob B. Simon\altaffilmark{1}, Xue-Ning Bai\altaffilmark{2,3,4}, James M. Stone\altaffilmark{2},
Philip J. Armitage\altaffilmark{1,5}, and Kris Beckwith\altaffilmark{1,6}}

\email{jbsimon@jila.colorado.edu}

\begin{abstract}
We use local numerical simulations to investigate the strength and nature of magnetohydrodynamic 
(MHD) turbulence in the outer regions of protoplanetary disks, where ambipolar diffusion is the 
dominant non-ideal MHD effect. The simulations include vertical stratification and assume zero net vertical
magnetic flux. We employ a 
super time-stepping technique to ameliorate the Courant restriction on the diffusive time step. 
We find that in idealized stratified simulations, with a spatially constant ambipolar Elsasser number Am, 
turbulence driven by the magnetorotational instability (MRI) behaves in a similar manner as in 
prior unstratified calculations. Turbulence dies away for ${\rm Am} \leq 1$, and becomes progressively 
more vigorous as ambipolar diffusion is decreased. Near-ideal MHD behavior is recovered for 
${\rm Am} \geq 10^3$. In the intermediate regime ($10 \leq {\rm Am} \leq 10^3$)  
ambipolar diffusion leads to substantial increases in both the period of the MRI dynamo cycle and the 
characteristic scales of magnetic field structures. To quantify the impact of ambipolar 
physics on disk accretion, we run simulations at 30~AU and 100~AU that include a vertical 
Am profile based upon far ultraviolet (FUV) ionized disk models. These models develop a 
vertically layered structure analogous to the Ohmic dead zone that is present at smaller radii. 
We find that, although the levels of surface turbulence can be strong (and consistent with constraints on 
turbulent line widths at these radii), the inferred accretion rates are at least an order of magnitude 
smaller than those observed in T~Tauri stars. This discrepancy is very likely due 
to the assumption of zero vertical magnetic field in our simulations and suggests that vertical
magnetic fields are essential for MRI-driven accretion in the outer regions of protoplanetary disks.
\end{abstract} 


\altaffiltext{1}{JILA, University of Colorado and NIST, 440 UCB, Boulder, CO 80309-0440}
\altaffiltext{2}{Department of Astrophysical Sciences, Princeton University, Princeton, NJ 08544}
\altaffiltext{3}{Hubble Fellow}
\altaffiltext{4}{Current address: Harvard-Smithsonian Center for Astrophysics, 60 Garden St.,
MS-51, Cambridge, MA 02138}
\altaffiltext{5}{Department of Astrophysical and Planetary Sciences, University of Colorado, Boulder, CO 80309}
\altaffiltext{6}{Tech-X Corporation, 5621 Arapahoe Ave., Suite A, Boulder, CO 80303}

\section{Introduction} 

The structure and evolution of protoplanetary disks play a crucial role in
the formation of stars and their planetary systems. Disk gas is observed
to  accrete onto the central star at rates that require  some form of
angular momentum transport substantially stronger than that provided by
molecular viscosity. Turbulence has long been suggested as the source of
enhanced transport  \citep{shakura73}. This turbulence  not only allows for
accretion, but can also play an important role in the formation and
subsequent evolution of planets. At early times, turbulence can act to
inhibit  dust settling and largely determine the collisional velocities
that affect the balance between fragmentation and coagulation  of these
particles \citep{ormel07,youdin07,birnstiel11}. Persistent pressure maxima  predicted by some
turbulence models \citep{barge95,johansen09,uribe11,simon12}  may act to concentrate particles, enhancing
their coagulation into larger particles. Once planetesimals have formed,
gravitational coupling  to turbulent fluctuations in the disk may affect
their growth \citep{ida08}.  Finally, the strength and nature of turbulence
determines whether the  critical co-orbital contribution to the Type~I
migration torque  remains unsaturated \citep{paardekooper11}.  

At a minimum, turbulence in protoplanetary disks will be generated
in  regions where the magnetorotational instability \cite[MRI;][]{balbus98} 
operates. Indeed, simulations of the non-linear evolution of the MRI under 
ideal magnetohydrodynamic (MHD) conditions yield sustained turbulence that
transports angular momentum outward at rates in general agreement with 
observations \citep{hartmann98a}. However, large regions of protoplanetary disks are expected to have
very low ionization fractions \cite[e.g.,][]{ilgner06}, which in turn result
in three  significant non-ideal MHD effects: Ohmic diffusion, ambipolar
diffusion, and the Hall effect \cite[see, e.g.,][]{armitage11}. The
relative importance of  these effects depends primarily upon the density
(as well as magnetic field strength).
Ohmic diffusion is efficient  at high densities, and is thus most
important  in the inner regions of the disk (outside a small zone very
close to the star where  thermal ionization of alkali metals provides
sufficient ionization throughout the disk column). At low gas densities,
such as in the outer disk, ambipolar diffusion becomes dominant, while at
intermediate densities, the Hall term is important \cite[e.g.,][]{kunz04}.

MRI physics and the phenomenological consequences of the non-ideal terms
have  been best-characterized in the case of Ohmic diffusion. The evolution
in this  limit depends upon the Elsasser number, defined as $\Lambda \equiv
v_{{\rm A},z}^2/\eta\Omega$, where $v_{{\rm A},z}$ is the $\alf$ velocity
in the vertical direction, $\eta$ is the Ohmic resistivity, and $\Omega$ is
the angular frequency of Keplerian rotation.  For $\Lambda$ less than order
unity, MRI turbulence is severely quenched  \citep{jin96,turner07}, while
for larger values the MRI saturation level will depend on the strength of
Ohmic diffusion; stronger diffusion leads to lower turbulence levels.
Combining these results with chemical models for  disks motivates the dead
zone model of disk accretion \citep{gammie96}. In this model, the disk is
well-ionized only in its surface layers due to  non-thermal sources
(X-rays, cosmic rays, and far ultraviolet (FUV) photons)  that penetrate
the disk from the exterior down to some column depth.  Closer to the
mid-plane, the ionization fraction is low, resulting in a small $\Lambda$,
and no MRI-driven turbulence
\citep[e.g.,][]{gammie96,fleming03,turner07,turner08,oishi09}. Thus,
MRI-driven accretion occurs in active layers only, leaving much of the disk
mass near the mid-plane magnetically inactive.

Qualitative changes to the predicted disk structure may equally result
from  ambipolar diffusion and the Hall effect, though for these terms the
understanding  of the non-linear behavior is incomplete.  The {\it linear} regime of the MRI
in the presence of the Hall term has been explored by \cite{wardle99}, \cite{balbus01},
and \cite{wardle12}.  A primary result from these studies is that the growth rate of the
MRI is strongly affected by the sign of ${\bmath \Omega} \cdot {\bmath B}$, i.e., how
the vertical magnetic field is aligned with the angular velocity vector.   The only study of
the non-linear, turbulent state of the MRI in the presence of the Hall term was carried
out in \cite{sano02a} and \cite{sano02b}.   Their numerical simulations included both
Ohmic diffusion {\it and} the Hall term, with the Ohmic contribution dominating significantly
over the Hall effect.  In this regime, the Hall term does not strongly influence the saturated
state of the MRI.  However, the regime in which the Hall term dominates has yet to be
explored through simulations \citep{wardle12}.  

Ambipolar diffusion arises from the imperfect coupling between ionized species and
neutrals. The linear analyses of \cite{blaes94}, \cite{kunz04}, and
\cite{desch04} showed that the growth of the MRI is damped when the collision
frequency between the neutrals and the ions is smaller than the
orbital frequency. This is intuitive;  neutrals need to communicate with the
ions faster than the timescale over which the MRI acts (i.e., the
dynamical one) in order for the neutrals to feel any MRI-like effect at
all. 

Initial two and three-dimensional simulations of non-linear MRI turbulence
in  the presence of ambipolar diffusion were carried out by \cite{maclow95}
and  \cite{brandenburg95}, respectively. These authors considered the
single-fluid,  ``strong-coupling" limit, valid when the recombination
timescale is much shorter than the dynamical time. This limit is generally 
applicable to protoplanetary disks \citep{bai11a,bai11b}. Their results 
agreed with the expectations of linear theory \citep{blaes94}; the MRI
only  operates if the collision frequency between neutrals and ions exceeds
the angular  frequency. Simulations in the alternate regime, where the
recombination timescale  for electrons is assumed to be very long, were
conducted by \cite{hawley98} using a two-fluid approach in which the ions
and neutrals were evolved separately, only interacting through collisions.
The primary result of this work was that for a collision frequency less
than $0.01\Omega$, the ions and neutrals behave independently, but for a
frequency larger than $100\Omega$, the gas behaves as though its fully
ionized. When both the collision and orbital frequencies are comparable,
the saturation level of the turbulence is primarily controlled by the ion
density.

The importance of ambipolar diffusion in the outer regions of protoplanetary  disks, the
advent of well-resolved mm-wave observations of these regions, and  advances
in numerical techniques, all motivate more detailed studies of how ambipolar
diffusion affects the saturated state of the MRI. \cite{bai11a}  carried out
shearing box simulations in the strong-coupling limit to determine
how the MRI saturation level correlates with dimensionless number, Am,
defined as the frequency for the neutrals to collide with the
ions divided by the orbital frequency
(see \S~\ref{num_method}). They found that for simulations with a net
toroidal magnetic flux, but no vertical magnetic flux, the turbulence dies
away for ${\rm Am} \leq1$, in line with previous studies.  For sustained
turbulence runs, the saturated turbulent stresses increase with increasing
Am, eventually asymptoting towards the ideal MHD level. However, in the
presence of a net vertical magnetic flux, turbulence can always be
sustained even for ${\rm Am} < 1$, assuming that the background vertical
magnetic flux is weak enough.  For low Am values, however, the resulting
turbulence levels are fairly small.

Following these lines of investigation, we study the effect of ambipolar
diffusion on the MRI in the outer region of protoplanetary disks by performing
numerical simulations using a more realistic disk structure than attempted
previously. We include vertical stratification (absent in all prior  work
except for that of \cite{brandenburg95}), which has been shown in some
 previous MRI calculations to lead to significant qualitative
changes in the non-linear evolution. For example, \cite{davis10} showed that
in the ideal MHD case, the turbulence properties converge with numerical resolution,
while in unstratified simulations (and for a vertical domain size of one scale height or
less, (Stone, private communication)), the stress level decreases with resolution with
no signs of convergence  \citep{fromang07a}. Similarly, \cite{simon11a} showed that
in the presence of Ohmic resistivity, vertical gravity can lead to large amplitude
fluctuations in the stress levels, a behavior that is absent without vertical gravity.

We also aim to translate our idealized understanding of the ambipolar-dominated MRI into
predictions for turbulence and accretion in the outer regions of protoplanetary disks.
We run simulations that mimic realistic conditions in the outer disk (which reflect the
chemistry calculations in \citet{bai11b} and the FUV ionization model of
\cite{perez-becker11b}). These simulations will provide a quantitative measure
of the turbulent saturation, structure, and evolution of the outer disk regions.

Our investigation is divided into two sets of studies, using different magnetic
configurations. The first, which we pursue here  (Paper~I), assumes zero net
vertical magnetic flux\footnote{Of  course due to the dynamo action of the MRI
\citep{davis10,simon11a}, the net radial and  toroidal fields are allowed to evolve
in time.}. Although it is unlikely that there will be exactly zero vertical magnetic flux penetrating
any given region of a disk, this is the most studied field configuration in
the literature (e.g., \citealp{stone96,fleming03}), and it allows us to make direct
comparisons between the ambipolar MRI and the previous simulations that include either Ohmic
resistivity or assume the gas is fully ionized. 
 It is also possible (though, not particularly likely)
that there will be some regions of protoplanetary disks that have negligible vertical magnetic fields; our results
will apply to such regions.
The second set of studies {\it will} include a non-zero vertical
net magnetic flux, and we defer that problem to Paper~II.

The structure of the paper is as follows.  In Section~\ref{method}, we
describe our equations, the methods used to solve them and the initial
conditions for our simulations. In Section~\ref{const_am}, we study the
effect of Am (here  assumed constant in space and time) on vertically
stratified MRI turbulence.   Then, in Section~\ref{variable_am}, we apply a realistic
protoplanetary disk model to allow for a spatially and temporally varying
Am. Section~\ref{implications} discusses the implications of our results
for real protoplanetary disks, and we wrap up with conclusions in
Section~\ref{conclusions}.

\section{Method}
\label{method}

\subsection{Numerical Method}
\label{num_method}

In this study, we use \textit{Athena}, a second-order accurate Godunov
flux-conservative code for solving the equations of MHD.  \textit{Athena}
uses the dimensionally unsplit corner transport upwind (CTU) method
of \cite{colella90} coupled with the third-order in space piecewise
parabolic method (PPM) of \cite{colella84} and a constrained transport
\citep[CT;][]{evans88} algorithm for preserving the $\del \cdot {\bmath
B}$~=~0 constraint.  We use the HLLD Riemann solver to calculate the
numerical fluxes \cite[]{miyoshi05,mignone07b}.  A detailed description
of the base \textit{Athena} algorithm and the results of various test problems
are given in \cite{gardiner05a}, \cite{gardiner08}, and \cite{stone08}.

Our set up is specialized and necessarily more complex than the base algorithm.  First, our
simulations utilize the shearing box approximation, which is a model
for a local, co-rotating disk patch whose size is small compared to the
radial distance from the central object, $R_0$.  This allows the construction of a local
Cartesian frame $(x,y,z)$ that is defined in terms of the disk's cylindrical co-ordinates $(R,\phi,z^\prime)$ via 
$x=(R-R_0)$, $y=R_0 \phi$, and $z = z^\prime$. The local patch 
co-rotates with an angular velocity $\Omega$ corresponding to
the orbital frequency at $R_0$, the center of the box; see \cite{hawley95a}. 
Thus, the equations to solve are:

\begin{equation}
\label{continuity_eqn}
\frac{\partial \rho}{\partial t} + \del \cdot (\rho {\bmath v}) = 0,
\end{equation}
\begin{equation}
\label{momentum_eqn}
\begin{split}
\frac{\partial \rho {\bm v}}{\partial t} + \del \cdot \left(\rho {\bm v}{\bm v} - {\bm B}{\bm B}\right) + \del \left(P + \frac{1}{2} B^2\right) \\
= 2 q \rho \Omega^2 {\bm x} - \rho \Omega^2 {\bm z} - 2 {\bm \Omega} \times \rho {\bm v} \\
\end{split}
\end{equation}
\begin{equation}
\label{induction_eqn}
\frac{\partial {\bmath B}}{\partial t} - \del \times \left({\bmath v} \times {\bmath B}\right) = \del \times \left[\frac{({\bmath J}\times {\bmath B})\times {\bmath B}}{\gamma \rho_i \rho}\right],
\end{equation}

\noindent 
where $\rho$ is the mass density, $\rho {\bmath v}$ is the momentum
density, ${\bmath B}$ is the magnetic field, $P$ is the gas pressure,
and $q$ is the shear parameter, defined as $q = -d$ln$\Omega/d$ln$R$.
We use $q = 3/2$, appropriate for a Keplerian disk.  We
assume an isothermal equation of state $P = \rho \cs^2$, where $\cs$
is the isothermal sound speed.  From left to right, the source terms
in equation~(\ref{momentum_eqn}) correspond to radial tidal forces
(gravity and centrifugal), vertical gravity, and the Coriolis force. The source
term in equation~(\ref{induction_eqn}) is the effect of ambipolar diffusion
on the magnetic field evolution, where $\rho_i$ is the ion density, and
$\gamma$ is the coefficient of momentum transfer for ion-neutral
collisions. Note that our system of units has the magnetic permeability $\mu = 1$, and
the current density is

\begin{equation}
\label{current}
{\bmath J} = \del \times {\bmath B}.
\end{equation}

Adopting this shearing box approximation allows for better resolution of small 
scales within the disk, at the expense of excluding global effects (those of 
scale $\sim R_0$). These scales could be physically important \citep{sorathia12,simon12}.    
However, the trade-off is worthwhile for our purposes, because we need to 
study not only models where ambipolar diffusion is dominant, but also situations 
where diffusion is {\it only} important on small scales.
 
The numerical integration of the shearing box equations require additions to
the \textit{Athena} algorithm, the  details of which can be found in \cite{stone10}
and the Appendix of \cite{simon11a}.  Briefly, Crank-Nicholson differencing is
used to conserve epicyclic motion exactly and orbital advection to subtract 
off the background shear flow \cite[]{stone10}. The $y$ boundary conditions
are strictly periodic, whereas the $x$ boundaries are shearing periodic
\cite[]{hawley95a}. The vertical boundaries are the outflow boundary
conditions described in \cite{simon11a}. The electromotive forces (EMFs) at
the radial boundaries are properly remapped to guarantee that the net
vertical magnetic flux is strictly conserved to machine precision using CT
\citep{stone10}.  In this paper, we only consider the case of zero net vertical magnetic
flux; thus, the methods we employ preserve this zero flux condition to machine accuracy.

The integration of the ambipolar diffusion term also requires some modifications
to the algorithm. Ambipolar diffusion is implemented in a first-order operator-split
manner as in \cite{bai11a}; the ambipolar diffusion term is integrated
separately from the ideal MHD integrator.  Furthermore, as is evident from
equation~(\ref{induction_eqn}), the ambipolar diffusion term
can be written as an EMF.  Thus integrating it is done via the CT method to
preserve $\del \cdot {\bmath B} = 0$.  The ambipolar diffusion EMFs are also
remapped at the radial boundaries in the same way as the ideal MHD EMFs in
order to maintain a zero vertical magnetic flux. In addition, before even
remapping these ambipolar diffusion EMFs at the radial boundaries,
we also remap the toroidal current densities $J_y$ (located at
cell edges) so that the line integral $\int J_y{\rm d}y$ along the inner
and outer radial boundaries are equal. We find
that this procedure is essential to avoid spurious numerical features at
shearing-box boundaries.\footnote{We note that strict magnetic flux conservation
(remap of the EMFs) was not enforced in \cite{bai11a}, in which case this additional
remap of $J_y$ was not necessary. Nevertheless, the variations in
vertical net magnetic flux in \cite{bai11a} simulations were tiny (less than $0.01\%$),
which did not affect their results.}

Throughout this paper, the strength of ambipolar diffusion will characterized by the
ambipolar Elsasser number
\begin{equation}
\label{am1}
{\rm Am}\equiv\frac{\gamma\rho_i}{\Omega},
\end{equation}

\noindent
which corresponds to the number of times a neutral molecule collides with the ions in a
dynamical time ($\Omega^{-1}$). Am can be rewritten as

\begin{equation}
\label{am2}
{\rm Am} = \frac{\va^2}{\eta_{\rm A} \Omega},
\end{equation}

\noindent
which is a form reminiscent of the Ohmic Elsasser number.  As shown by equation~(\ref{am1}),
Am is independent of the $\alf$ speed; this comes about because the ambipolar diffusivity,
$\eta_{\rm A}$ is defined as 
\begin{equation}
\eta_{\rm A}\equiv\frac{\va^2}{\gamma \rho_i}.
\end{equation}

This diffusivity is responsible for determining the diffusive time step in a Courant limited calculation;
$\Delta t_{\rm diff} \propto \Delta x^2/\eta_{\rm A}$.  Since the diffusivity is proportional to the $\alf$
speed squared, it can become very large in the upper disk regions, making the Courant limited time step
extremely small in some of our calculations.

To circumvent this issue, we have implemented the super time-stepping (STS) technique of
\cite{alexiades96} to accelerate our calculations. The STS technique has already been
successfully implemented and tested for studying ambipolar diffusion in multi-fluid codes by
\citet{osullivan06} and \citet{osullivan07} and in a single-fluid code by
\citet{choi09}. Our implementation is similar to theirs, as we describe in detail in the Appendix.

\subsection{Am Profiles}

For most of our simulations, we fix Am to be constant in order to study the effect
of ambipolar diffusion on the non-linear saturation of the MRI in the presence of
vertical stratification.  This is designed to be the next logical step in extending the
work of  \citet{bai11a} where vertical stratification was not included. We have
considered Am=$1, 10, 100, 300, 10^3$ and $10^4$. Although this prescription of a
constant Am profile is highly simplified, it is a necessary, incremental step between
the constant Am models without vertical gravity \citep{bai11a} and simulations that
incorporate a more realistic prescription for ambipolar diffusion, which we also carry
out (see below). 

The results of \cite{bai11b} show that the physical value of Am in the outer regions of PPDs is of order unity or less.
Recently, \citet{perez-becker11b} pointed out that the surface layer of
protoplanetary disks should be much better ionized due to far ultraviolet (FUV) photon
ionization from the central star; these photons almost fully ionize the carbon and sulfur to
overcome the effects of recombination onto dust grains. Their results imply
 a large ionization fraction ($f\sim10^{-5}$) down to a small penetration
depth of $\Sigma_{\rm FUV}\sim0.01-0.1$ g cm$^{-2}$, relatively independent of disk radius.
Such an ionization fraction should significantly reduce the strength of ambipolar diffusion (i.e.,
increase Am) in the disk surface layers.

Thus, in our second set of simulations, we include the effect of FUV ionization at
the disk surface layers to give Am a more realistic spatial and temporal dependence.
Since we are not including Ohmic resistivity \citep{gammie96} in our calculations, these
particular models are only appropriate for the outer regions of protoplanetary disks (e.g.,
beyond $\sim 10$ AU) where ambipolar diffusion dominates Ohmic diffusion \citep{kunz04,armitage11}.
 We adopt a minimum-mass solar nebular disk model with surface density of
$\Sigma=1700R_{\rm AU}^{-3/2}$g cm$^{-2}$ \citep{weidenschilling77,hayashi81}, where
$R_{\rm AU}$ is the disk radius measured in AU.  We can express the value of Am within
the FUV ionized layer as follows \citep{bai12}

\begin{equation}
\label{Am_FUV}
{\rm Am_{\rm FUV}} \approx3.3\times10^7
\bigg(\frac{f}{10^{-5}}\bigg)\bigg(\frac{\rho}{\rho_{0,{\rm mid}}}\bigg)R_{\rm AU}^{-5/4}\ ,
\end{equation}

\noindent
where $f$ is the ionization fraction and $\rho_{0, {\rm mid}}$ is the midplane density.
For simplicity, we fix $f=10^{-5}$, $\rho_{0, {\rm mid}} = 1$, and assume a penetration
depth of $\Sigma_{\rm FUV}=0.1$g cm$^{-2}$ (which is slightly different from that in
\citealp{bai12}). We conduct two simulations that correspond to radial
locations at $R=30$ AU and $R=100$ AU. Assuming the density profile is
Gaussian (see Equation (\ref{density_init})), one finds that the base of the FUV layer
(at which the column density equals $\Sigma_{\rm FUV}$) is located at
$z_b=1.7H$ for $R=30$ AU and $z_b=1.1H$ for $R=100$ AU ($H$ is the vertical scale height as
defined in equation~(\ref{scale_height}) below). In our simulations,
we set Am=1 for $-z_b<z<z_b$ as a proxy based on the calculations of
\cite{bai11b}, and use Equation (\ref{Am_FUV}) for the ionized surface layers of the disk.  In principle, 
Am $< 1$ with the inclusion of grains \citep{bai11b}.  However, as we will see, for Am = 1 ambipolar diffusion is
sufficiently strong to quench the MRI in this mid-plane region; thus, going to lower
values of Am is unnecessary for the purposes of this study.
Finally, we keep the value of $z_b$ fixed throughout the calculation for simplicity.

From these considerations, the value of Am changes quite dramatically from Am=1
to Am$\approx3\times10^4$ at the base of the FUV layer. This very large transition
is smoothed over roughly 7 grid zones so as to prevent a discontinuous transition in Am.
The smoothing functions we apply are based upon the error function (ERF).   Thus, the
complete profile of Am for these runs is given by

\begin{equation}
\label{amc}
\small
{\rm Am} \equiv \left\{ \begin{array}{ll}
 {\rm Am_{\rm FUV}} & \quad 
\mbox{$z \ge z_b + \Delta z$} \\
1 + \frac{1}{2}{\rm Am_{\rm FUV}}S^+(z)  & \quad
\mbox{$z_b - n\Delta z < z < z_b + \Delta z$} \\
1 & \quad
 \mbox{$-z_b+n\Delta z \le z \le z_b - n \Delta z$} \\
1 + \frac{1}{2}{\rm Am_{\rm FUV}}S^-(z) & \quad
\mbox{$-z_b-\Delta z < z < -z_b + n\Delta z$} \\
{\rm Am_{\rm FUV}} & \quad
\mbox{$z \le -z_b - \Delta z$}
\end{array} \right.
\end{equation}

\noindent
where $S^+(z)$ and $S^-(z)$ are the smoothing functions defined as 
 
\begin{equation}
\small
\label{splus}
S^+(z) \equiv 1+{\rm ERF}\left(\frac{z-0.9z_b}{\Delta z}\right),
\end{equation}
\begin{equation}
\small
\label{sminus}
S^-(z) \equiv 1-{\rm ERF}\left(\frac{z+0.9z_b}{\Delta z}\right),
\end{equation}

\begin{figure}
\begin{center}
\includegraphics[width=0.5\textwidth,angle=0]{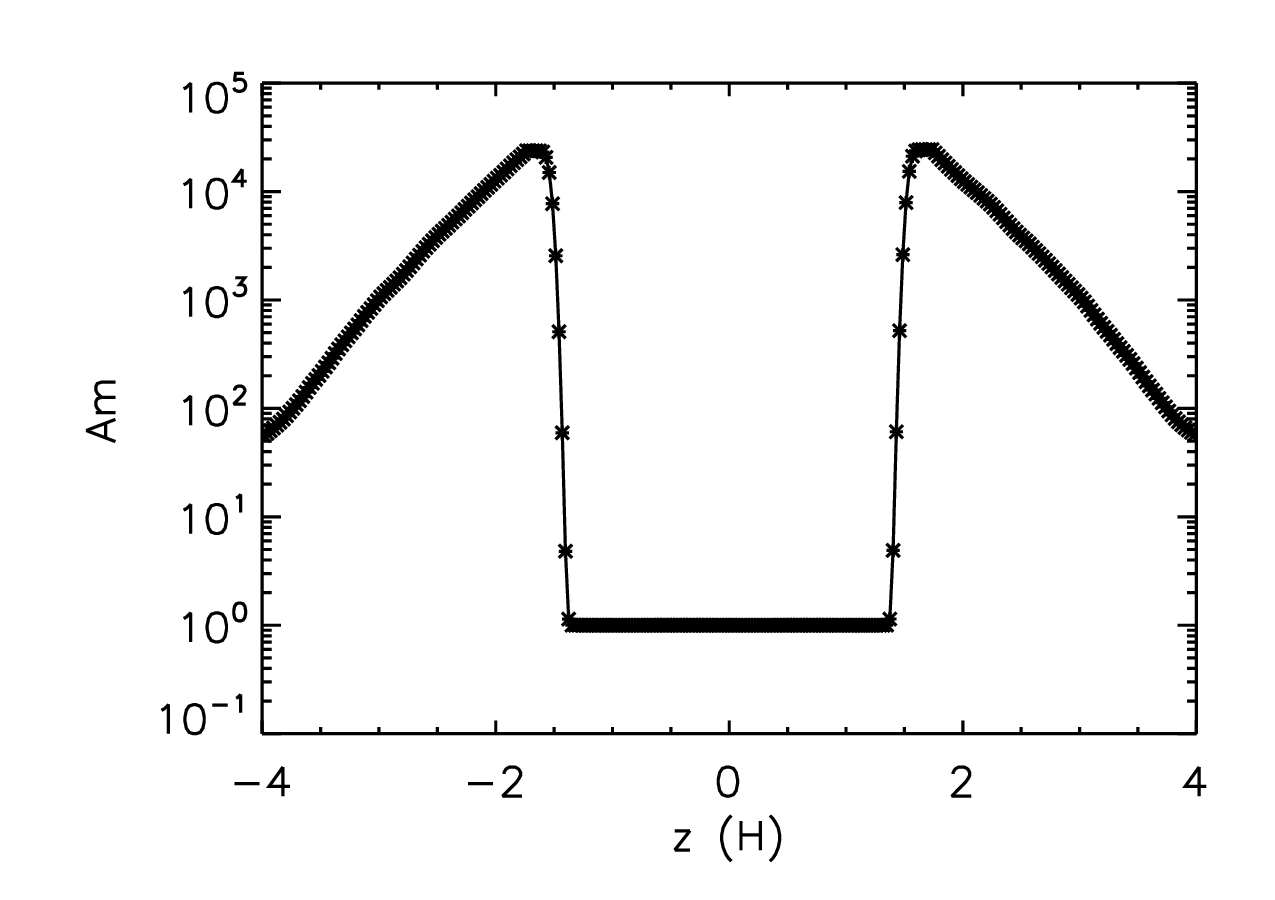}
\end{center}
\caption{
Vertical profile for Am at $R_{\rm AU} = 30$. The profile corresponds to the initial time of Z30AU, which is orbit 22 from the Am = $10^5$ run with the same domain size.  The value of Am has been averaged horizontally.  The units of the $x$ axis are the vertical scale height, $H$.
The asterisks denote the locations of grid zones.  Am transitions from Am = 1 to
Am = ${\rm Am_{\rm FUV}}$ (as defined in the text), and this transition
is smoothed over roughly 7 grid zones using the error function.
}
\label{amz}
\end{figure}

\noindent
Here, $n = 8$ and $\Delta z = 0.05 H$.  These numbers were chosen to give a reasonably
resolved transition region between Am = 1 and ${\rm Am_{\rm FUV}}$.  For a visual
representation of the rather complex equation~(\ref{amc}), we plot in Fig.~\ref{amz} the
vertical profile of Am (averaged over $x$ and $y$) for the run
at $R_{\rm AU} = 30$ at the initial time, referring here to when the run was restarted from C1e5 (see below).
The asterisks denote the grid cell centers; as previously mentioned,
the transition region is resolved by $\sim 7$ zones.

\subsection{Simulations}

We have run simulations with several domain sizes and Am profiles.  
All of the simulations with Am $< 10^5$ or with spatially and temporally varying Am are
initialized from the turbulent state of  a ``starter" calculation with the same domain size
but with Am = $10^5$ (i.e., reasonably close to ideal MHD). 

These starter simulations are initialized with a density 
corresponding to isothermal hydrostatic equilibrium.

\begin{equation}
\label{density_init}
\rho(x,y,z) = \rho_0 {\rm exp}\left(-\frac{z^2}{H^2}\right),
\end{equation}

\noindent where $\rho_0 = 1$ is the mid-plane density, and $H$
is the scale height in the disk,

\begin{equation}
\label{scale_height}
H = \frac{\sqrt{2} \cs}{\Omega}.
\end{equation}

\noindent The isothermal sound speed, $\cs = 7.07 \times 10^{-4}$,
corresponding to an initial value for the mid-plane gas pressure of $P_0 = 5
\times 10^{-7}$.  With $\Omega = 0.001$, the value for the scale height
is $H = 1$.   A density floor of $10^{-4}$ is applied to the physical domain as
too small a density leads to a large $\alf$ speed and a very small
time step.  Furthermore, numerical errors make it difficult
to evolve regions of very small plasma $\beta$ (ratio of thermal pressure to
magnetic pressure).  

The initial magnetic field is purely toroidal and has a constant $\beta = 100$ throughout
the domain (thus, $B_y^2$ has a Gaussian shape like the density). Random
perturbations are added to the density and velocity components to seed the MRI.

The remaining simulations are restarted from orbit 50 (orbit 22 for the simulations with
domain size $8H\times16H\times8H$\footnote{We choose a different restart time for
these calculations because we decided midway through running our simulations that a
larger number of cores is significantly more efficient for the variable Am runs.  Thus, we
had to redo the Am = $10^5$ calculations, and orbit 22 was chosen because the
density-weighted stress at this time was roughly equal to orbit 50 of the lower core
version of this calculation.}) of their corresponding domain size simulation with
Am = $10^5$.  They are listed in Table~\ref{tbl:sims}. The label for each calculation
describes whether the value of Am is constant with height, labeled C, or varies according
to equation~(\ref{amc}), labeled Z.  For the constant Am simulations, the number following
the C is the value of Am.  For the spatially varying Am calculations, the number afterwards
(along with the AU) describes the radial location of the shearing box in our protoplanetary
disk model in units of AU.  An S (L) following the Am value corresponds to a domain size of
$2H \times 4H \times 8H$ ($8H \times 16H \times 8H$), which is smaller (larger) than the
$4H \times 8H \times 8H$ size of most of the constant Am calculations. The ``starter"
simulation for the $4H\times8H\times8H$ runs is also included in the table, labeled C1e5.
Finally, all of our calculations are carried out at a resolution of 36 grid zones per $H$.
 
\section{Results}
\label{results}

\subsection{Constant Am Calculations}
\label{const_am}

\begin{figure}
\begin{center}
\includegraphics[width=0.5\textwidth,angle=0]{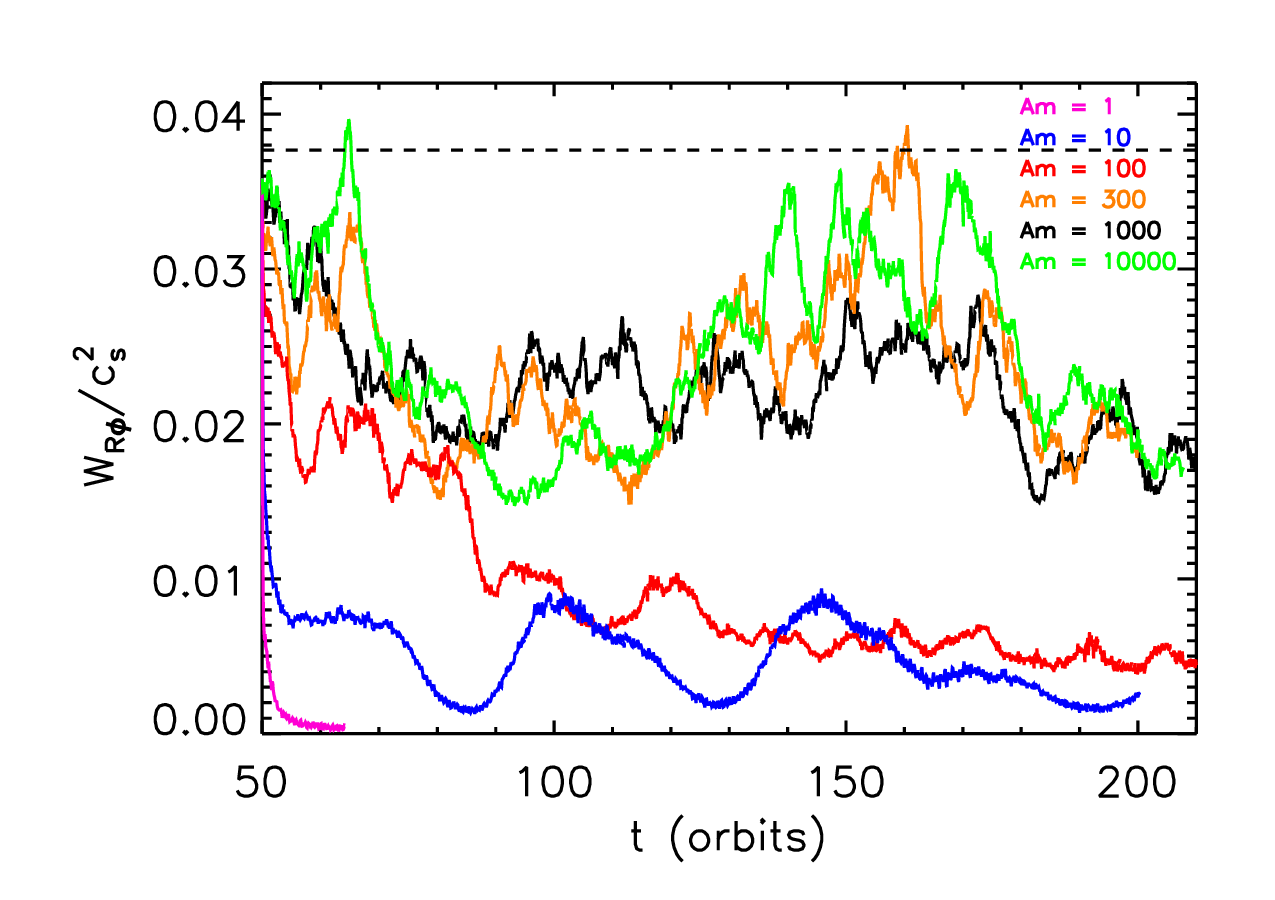}
\end{center}
\caption{
Density-weighted volume average of the total (Maxwell and Reynolds) stress, normalized by the square of the sound speed, versus time for the
standard $4H\times8H\times8H$ simulations.  The magenta line corresponds to Am = 1, blue
to Am = 10, red to Am = 100, orange to Am = 300, black to Am = 1000, and green to Am = $10^4$.
The horizontal dashed line corresponds to the time-averaged (from orbit 25 to 53) stress value for
Am = $10^5$ from which the other runs were initialized.  After an initial transient of $\sim 50$ orbits,
the simulations with Am $\gtrsim 10$ are sustained.  There is a general trend of increasing stress
level with increasing Am.  The Am = 1 case has turbulence that decays away rapidly.
}
\label{hist_4x8x8}
\end{figure}

We begin by applying some standard diagnostics to the set of calculations with
constant values of Am.  The first such diagnostic is the density-weighted Maxwell
and Reynolds stresses \cite[see equation (37) of][]{balbus98}, defined as 

\begin{equation}
\label{wrp}
W_{R\phi} = \frac{\left\langle \rho v_x \delta v_y
- B_xB_y\right\rangle}{\left\langle \rho\right\rangle}
\end{equation}

\noindent
where the angled brackets denote a volume average over the whole domain.  
Figure~\ref{hist_4x8x8} shows the time evolution of this total stress for the $4H\times8H\times8H$ runs, normalized by the square
of the sound speed.   The dashed line indicates the
averaged value (from orbit 25 to 53) of the run with Am = $10^5$ from which all of the other
$4H\times8H\times8H$ runs were restarted.  The different Am values are denoted by the color.
The runs with ${\rm Am} > 1$ appear to adjust on a roughly 50 orbit timescale after which a
statistical steady state follows.  In general, the stress increases with increasing Am (decreasing
diffusion) for these runs.  However, the Am = 10 and Am = 100 runs have roughly the same
values at late times, as do the Am = 300, 1000, and 10000 runs.

The Am = 1 case deserves extra attention.  From Fig.~\ref{hist_4x8x8}, it would appear that the
turbulence completely dies away.  A closer examination of the stress histories show that the
 Maxwell stress levels out to a small, but positive value, while continuing to slowly decrease with time.  The 
Reynolds stress approaches an oscillatory behavior which occasionally brings it below zero. 
Space-time plots of various quantities in this run indicate that the gas is no longer MRI turbulent.
The remnant Maxwell stress is the result of a residual large scale $B_x$ and $B_y$ field near the
mid-plane, and the Reynolds stress appears to arise from residual waves propagating through the
box.  The longer term behavior of this Am value could not be examined because even with STS,
the diffusion limited time step is very small for Am = 1; running it out further would be very
computationally expensive.  However, these results strongly indicate that the MRI turbulence has completely
decayed away, consistent with the results of \cite{bai11a}. This behavior will play an
important role in the variable Am simulations of Section~\ref{variable_am}.   

\begin{figure}
\begin{center}
\includegraphics[width=0.5\textwidth,angle=0]{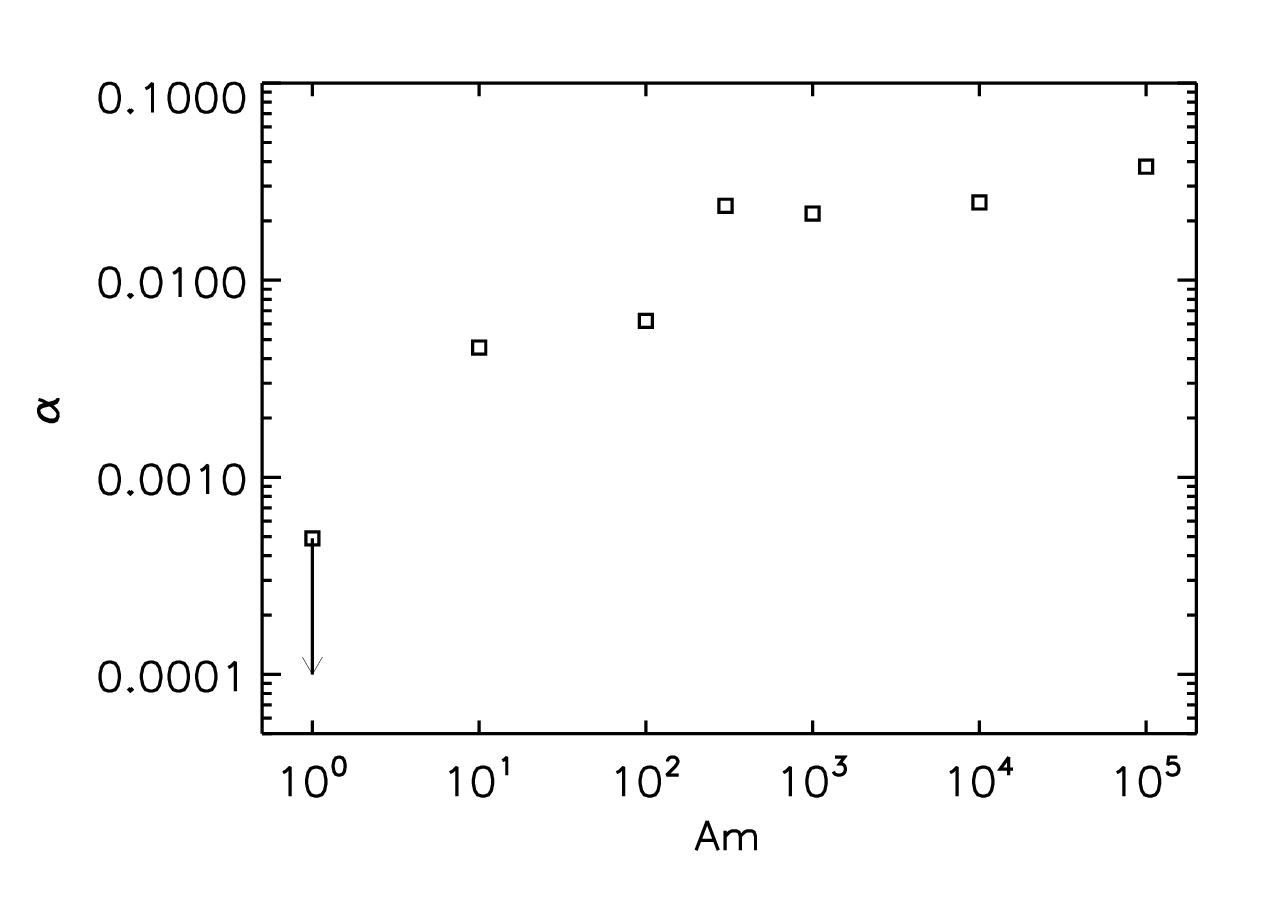}
\end{center}
\caption{
Time-averaged total stress (i.e., $\alpha$) as a function of Am for the standard $4H\times8H\times8H$ simulations.  There is a general trend of increasing stress level with increasing Am. 
}
\label{alpha_4x8x8_log}
\end{figure}

We time-average this normalized stress and define the Shakura-Sunyaev $\alpha$ parameter,

\begin{equation}
\label{alpha}
\alpha = \overline{\frac{W_{R\phi}}{\cs^2}}
\end{equation}

\noindent
where the overbar denotes the time average, which is done from orbit 100 onwards for most of
the constant Am runs with Am $> 1$; from orbit 72 onwards for runs C10L, Z30AU, and Z100AU; and
from orbit 25 to 53 for the Am = $10^5$ ``starter" simulation.  Fig.~\ref{alpha_4x8x8_log} displays $\alpha$ versus Am for these runs.  The arrow on
the Am = 1 run indicates that that the stress level is continually decreasing.  The trend of $\alpha$
with Am can be compared with the unstratified simulations shown in Fig. 10 of \cite{bai11a}. These
trends roughly agree, though the results from \cite{bai11a} show a monotonic increase of $\alpha$
with Am, whereas our results show that different Am values can lead to very similar $\alpha$ values.

\begin{deluxetable}{l|ccc}
\tabletypesize{\small}
\tablewidth{0pc}
\tablecaption{Shearing Box Simulations\label{tbl:sims}}
\tablehead{
\colhead{Label}&
\colhead{Ambipolar Diffusion}&
\colhead{Domain Size}&
\colhead{$\alpha$} \\
\colhead{ }&
\colhead{ }&
\colhead{$(L_x \times L_y \times L_z) H$}&
\colhead{ } } 
\startdata
C1 & Am = 1, constant & $4 \times 8 \times 8$ & decayed \\
C10 & Am = 10, constant & $4 \times8 \times 8$ & 0.0046\\
C10L & Am = 10, constant & $8 \times 16 \times 8$ &  0.0055\\
C100S & Am = 100, constant & $2 \times 4 \times 8$ & 0.070\\
C100 & Am = 100, constant & $4 \times8 \times 8$ & 0.0062 \\
C300 & Am = 300, constant  & $4 \times8 \times 8$ & 0.024\\
C1000 & Am = 1000, constant  & $4 \times8 \times 8$ & 0.022\\
C10000 & Am = $10^4$, constant & $4 \times8 \times 8$ & 0.025\\
C1e5 & Am = $10^5$, constant & $4 \times8 \times 8$ & 0.038\\
Z30AU & Am at $R = 30$AU & $8 \times 16 \times 8$ & 0.0016  \\
Z100AU & Am at $R = 100$AU& $8 \times 16 \times 8$ &  0.0015 \\
\enddata
\end{deluxetable}

\begin{figure*}
\begin{center}
\includegraphics[width=1\textwidth,angle=0]{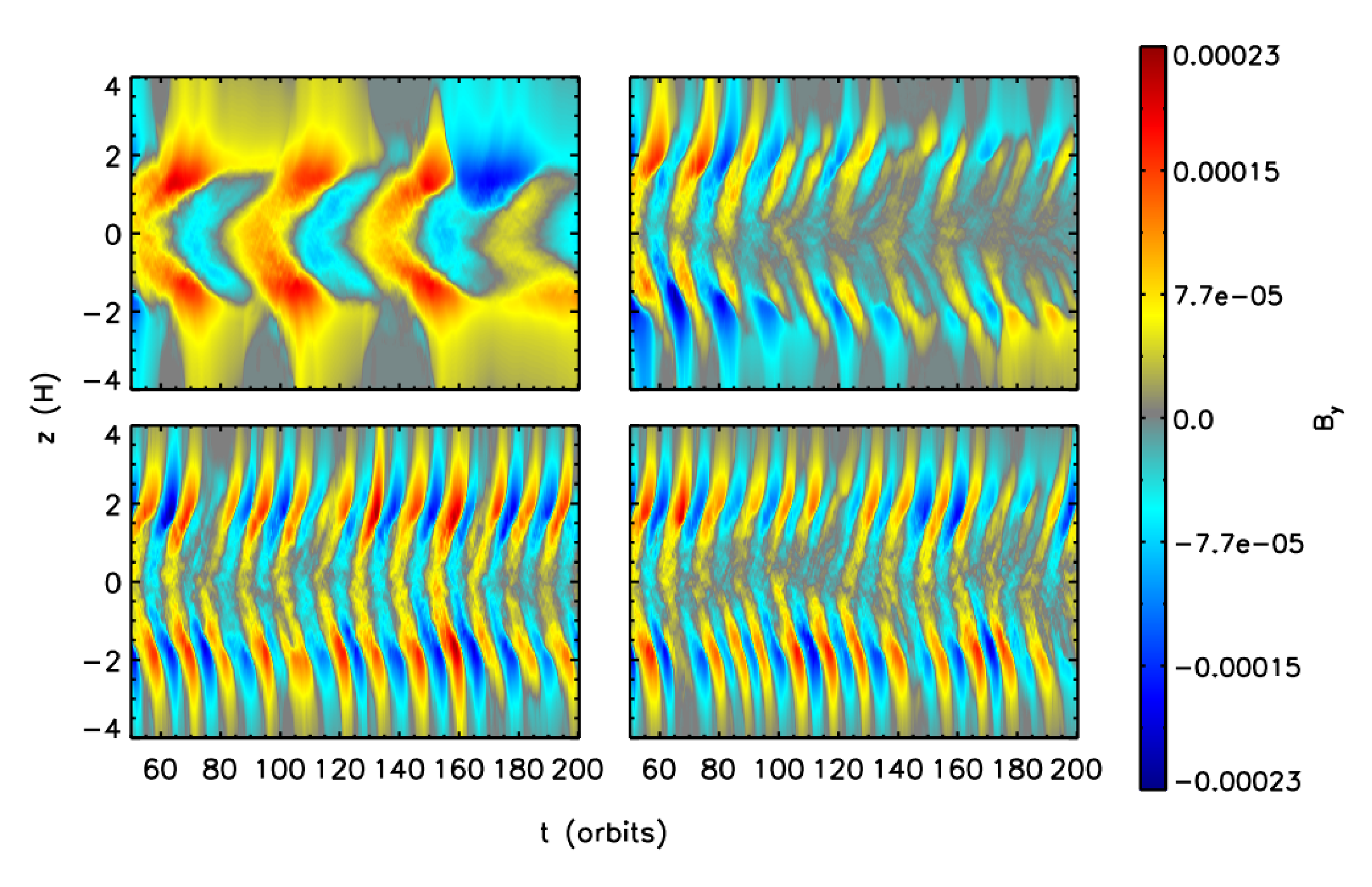}
\end{center}
\caption{Space-time plot of $B_y$ averaged over $x$ and $y$ for Am = 10 (upper left),
Am = 100 (upper right), Am = 300 (lower left), and Am = 1000 (lower right).  The
``butterfly" dynamo is present in all simulations, but the period of the $B_y$ flipping
increases with decreasing Am.  In particular, the period is $\sim 40-50$ orbits for
Am = 10 and $\sim 15-20$ orbits for Am = 100.  For the other two cases, the period is
$\sim 10$ orbits, equal to that in ideal MHD calculations.
}
\label{sttz_by}
\end{figure*}

This difference may be attributable to different background magnetic field strengths as the background
field evolves via the usual MRI dynamo \cite[e.g.,][]{davis10,simon11a}. To
understand this further, first let us consider the space time diagrams of the toroidal field $B_y$ for C10,
C100, C300, and C1000 as shown in Fig.~\ref{sttz_by}.  In these diagrams, the field has been
averaged over $x$ and $y$ and is plotted in the $(t,z)$ plane.  The most obvious feature from these
diagrams is that the period of the dynamo flipping of $B_y$ changes with Am; as ambipolar diffusion
becomes stronger, the period increases.  In particular, for Am = 10, the period is $\sim 50$ orbits,
and for Am = 100 (only considering times past the initial 50 orbit transient period), the period is
$\sim 15-20$ orbits.  For Am $\ge 300$, the period is $\sim 10$ orbits as is usually observed in
stratified MRI simulations.

\begin{figure}
\begin{minipage}[!ht]{8cm}
\begin{center}
\includegraphics[width=1\textwidth,angle=0]{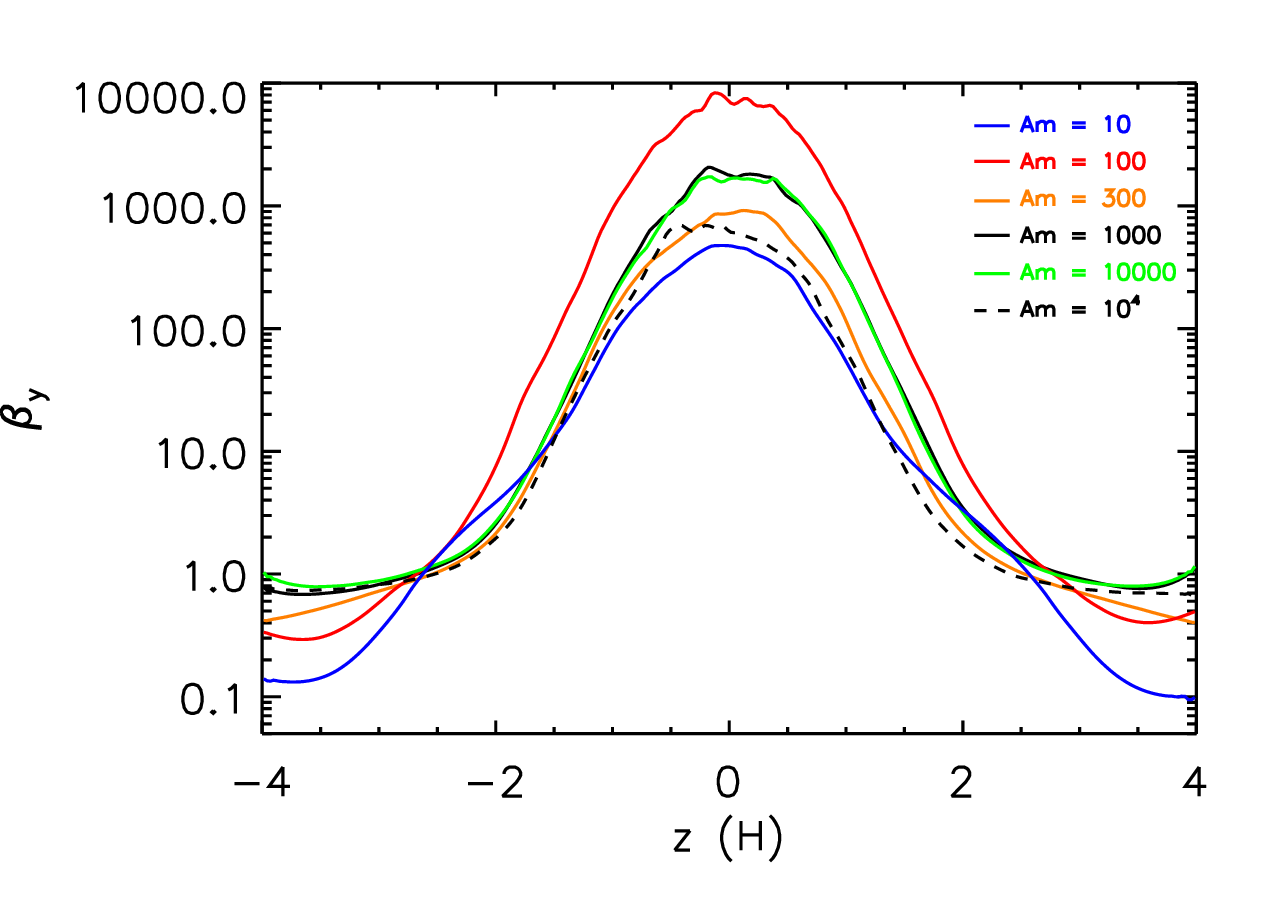}
\end{center}
\end{minipage}
\begin{minipage}[!ht]{8cm}
\begin{center}
\includegraphics[width=1\textwidth,angle=0]{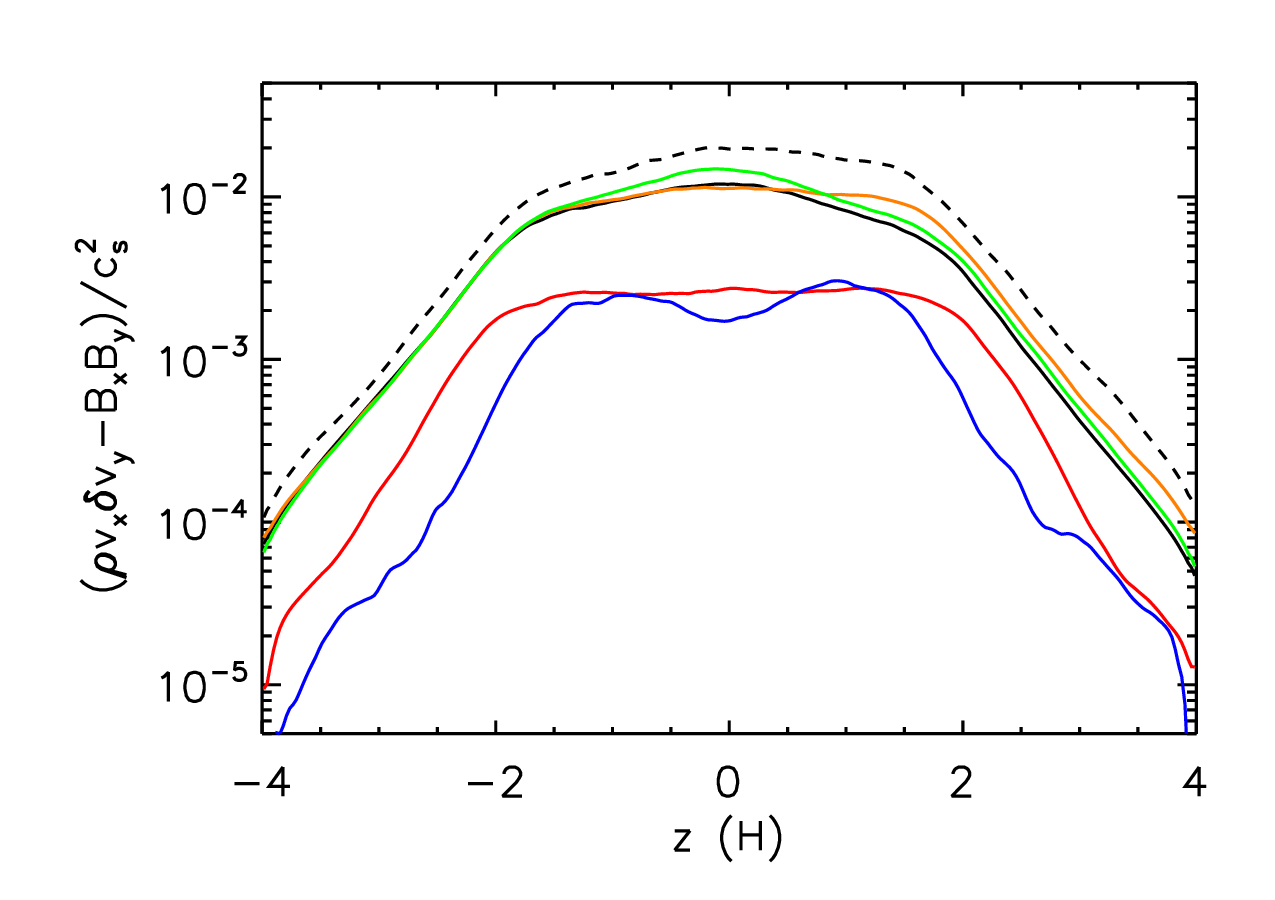}
\end{center}
\end{minipage}
\caption{Vertical profiles of $\beta_y$ as defined by equation~(\ref{betay}) (top) and the
total stress normalized by the square of the sound speed (bottom).  The quantities have
been averaged over $x$ and $y$ and over time from orbit 100 onwards as described in
the text.  The blue line corresponds to Am = 10, red to Am = 100, orange to Am = 300,
black to Am = 1000, green to Am = $10^4$, and dashed line to Am = $10^5$ (the time
average for this run is done from orbit 25 to the end of the calculation).  The stress
increases with Am roughly uniformly at all heights.  There is no obvious trend between
$\beta_y$ and Am.
}
\label{vert_4x8x8}
\end{figure}

The most relevant feature here, however is that the background toroidal field strength is
different for different values of Am. In \cite{bai11a}, it was found that with zero net vertical flux,
the stress level increases with increasing net toroidal flux (which is mostly conserved in
unstratified simulations). Therefore, in our stratified simulations, {\it two} effects are expected
to determine the saturated stress values: the value of Am and the background toroidal field
strength, set by the dynamo.
To demonstrate this effect more robustly, we calculate a version
of the plasma $\beta$ for the background toroidal field,
\begin{equation}
\label{betay}
\beta_y \equiv 2\overline{\langle P\rangle}/\overline{\langle B_y\rangle^2},
\end{equation}
\noindent
where the overbars indicate a time average (from orbit 100 onwards) and the angled
brackets denote an average over $x$ and $y$.  This quantity is representative of the
amplitude of the oscillating background toroidal field.   $\beta_y$ is a function of $z$
only, and we plot it along with the vertical profile of the total stress (which has again,
been averaged in time and for all $x$ and $y$)  in Fig.~\ref{vert_4x8x8}.

The stress profile reveals the same behavior as that in Fig.~\ref{alpha_4x8x8_log}; there
is a general trend of increasing stress with increasing Am. Furthermore, this increase
occurs uniformly across all $z$.  However, C300 and C1000 have roughly the same stress
profiles, and C10 peaks at around the same value as C100.  Examining the $\beta_y$ for
these particular simulations, we see that C300 has a lower value (stronger field) than does
C1000.  Similarly, C10 has a significantly smaller $\beta_y$ than C100. These results
confirm that it is indeed the larger background toroidal field strength that make the stress
levels in run C10 approach that in run C100, and the stress in run C300 approach that
in run C1000. We note, however, that C1000 and C10000
have both the same $\beta_y$ profiles and the same stress profiles.  This could indicate that
for Am $> 1000$, the turbulence levels are approaching that of ideal MHD.  The slightly higher
stress for C1e5 would then be explained by its lower $\beta_y$.  

It remains unclear why the background field strength varies in the way that it does.  Could
this also be controlled by the value of Am?  This is not unreasonable considering that
ambipolar diffusion already affects the period of the toroidal field flipping.  The question of
exactly how Am and the dynamo are related is very open.  Unfortunately, exploring it in
detail would take us too far from our goals in this paper, and so, we leave it for future work.

The final diagnostic we employ is the two-point autocorrelation function first used in the
context of MRI simulations in \cite{guan09a}.  We employ this diagnostic for similar reasons
as those in \cite{simon12}; we wish to determine the structure of the turbulent magnetic
field and check that the domain sizes we use are sufficiently large to properly capture
important turbulent scales.  Thus, we define the autocorrelation function of the $i^{\rm th}$
component of the {\it perturbed} magnetic field as

\begin{equation}
\label{acf}
{\rm ACF}(\delta B_i(\bmath{\Delta x})) = \overline{\frac{\int \delta B_i(t,\bmath x)
\delta B_i(t,\bmath x + \bmath{\Delta x}) d^3{\bmath x}}{\int \delta B_i(t,\bmath x)^2 d^3{\bmath x}}},
\end{equation}

\noindent
where $\delta B_i$ is the value of $B_i$ after subtracting off the horizontal mean field.  In equation form,

\begin{equation}
\label{deltab}
\delta B_i(x,y,z) \equiv B_i(x,y,z)-\langle B_i\rangle_{xy}(z),
\end{equation}

\noindent
and the average denoted by $\langle\rangle_{xy}$ is the horizontal average.
We have defined the ACF to be normalized by its maximum value (at
$\Delta x = \Delta y = \Delta z = 0$).  The ACF of the total turbulent magnetic field is then
defined as ACF($\delta B$) = ACF($\delta B_x$) + ACF($\delta B_y$) + ACF($\delta B_z$).
The overbar denotes a time average done from orbit 100 to 125 in all cases.

\begin{figure*}
\begin{center}
\includegraphics[width=1\textwidth,angle=0]{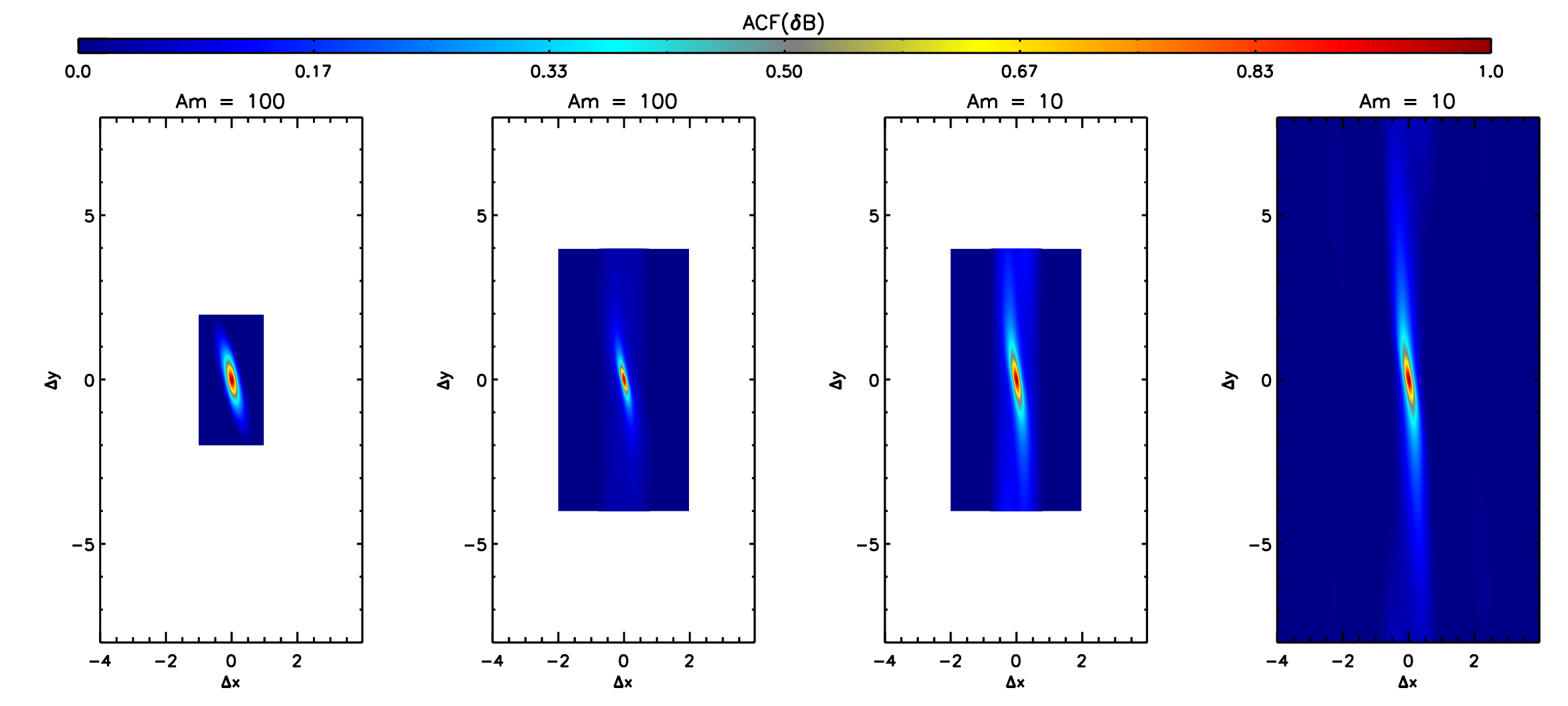}
\end{center}
\caption{Autocorrelation function (ACF) of the magnetic field, as defined by
equation~(\ref{acf}), for simulations (from left to right) with Am = 100 and size
$2H\times4H\times8H$, Am = 100 and size $4H\times8H\times8H$, Am = 10 and size
$4H\times8H\times8H$, and Am = 10 and size $8H\times16H\times8H$.  As Am is
decreased, larger and larger domains are needed to properly contain the ACF.  
Furthermore, the tilted centroid feature becomes less tilted with respect to the $y$ axis and more elongated as Am is decreased. 
}
\label{corr2d_b}
\end{figure*}

From the figure, it appears that the Am = 100 ACF is roughly consistent between a
domain size of $2H\times4H\times8H$ and $4H\times8H\times8H$, though there is a
slight difference in the size of the tilted centroid.  However, as we will see shortly,
$2H\times4H\times8H$ is actually too small for Am = 100.  The standard box size,
$4H\times8H\times8H$, appears to properly contain the ACF for Am = 100, but not as
well for Am = 10.  The centroid of the Am = 10 case is larger and appears to have a
longer tail that intersects the toroidal boundary.  Going to an even larger domain,
$8H\times16H\times8H$, the ACF($B$) for Am = 10 looks more well contained, though
the very end of the tail does appear to touch the toroidal boundary.  This effect is not
as dramatic as in the smaller domain.  Going to an even larger domain and running
Am = 10 is prohibitively expensive given our current computational resources.

\begin{figure}
\begin{center}
\includegraphics[width=0.5\textwidth,angle=0]{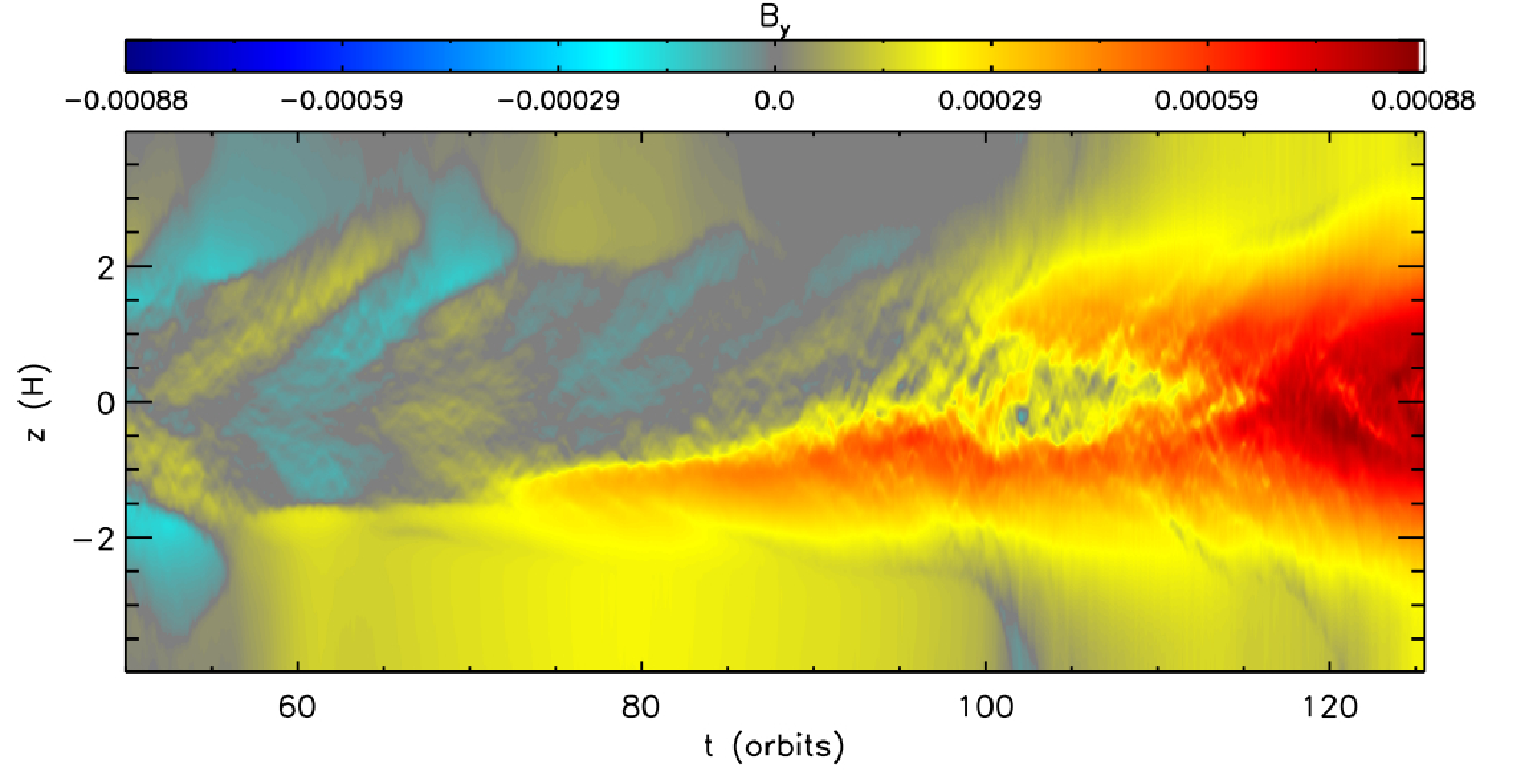}
\end{center}
\caption{Space-time plot of $B_y$ averaged over $x$ and $y$ for the Am = 100 run at domain size $2H\times4H\times8H$.  There is a remnant dynamo behavior that rapidly dies away as the simulation adjusts to the new value of Am.  Eventually, the dynamo activity ceases altogether, which is inconsistent with the larger domain counterpart of Am = 100.  This domain size appears to be too small to properly capture the dynamo at Am = 100.}
\label{sttz_by_am100_2x4x8}
\end{figure}

Returning to the smallest domain, we note some odd features.  Despite the reasonable
ACF, an inspection of the stress history and the $\alpha$ value (see Table~\ref{tbl:sims})
show this calculation to be quite different than Am = 100 at larger domain sizes.  An
examination of the space-time evolution of $B_y$, Fig.~\ref{sttz_by_am100_2x4x8},
brings the point home further, as it indicates that the dynamo behavior is completely shut
off for this particular run.  This is again inconsistent
with the larger domain Am = 100.  Thus, we conclude that $2H\times4H\times8H$ is too
small of a domain for Am = 100 and will likely be too small for smaller values of Am as well.

\begin{figure}
\begin{center}
\includegraphics[width=0.5\textwidth,angle=0]{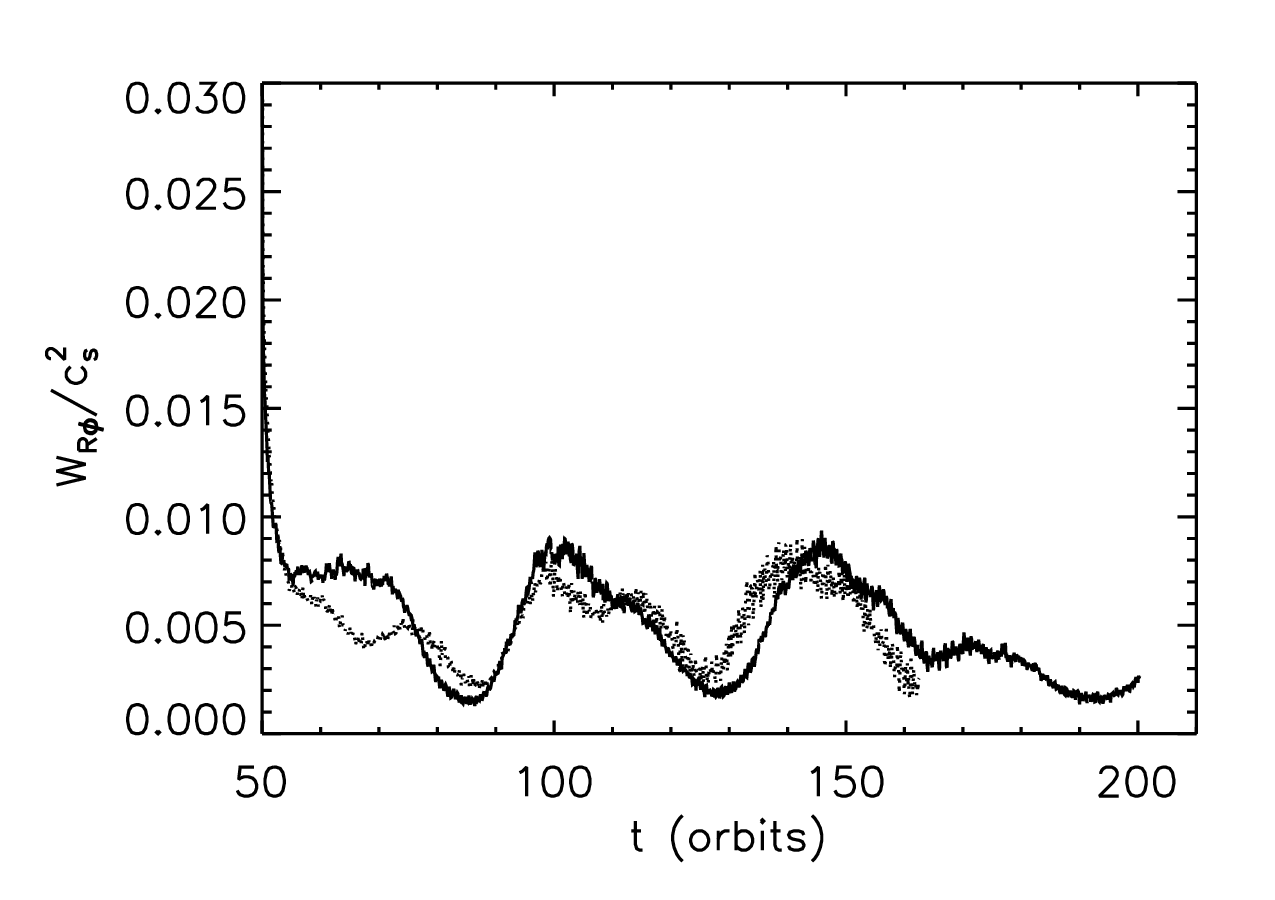}
\end{center}
\caption{Density-weighted volume average of the total (Maxwell and Reynolds) stress, normalized by the square of the sound speed, versus time for the Am = 10 simulation at a domain size of $4H\times8H\times8H$ (solid line) and $8H\times16H\times8H$ (dotted line).  Note that the $8H\times16H\times8H$ run was restarted from orbit 22 of a different ``starter" simulation.  For comparison purposes, we translated the solution to the right by 28 orbits. The curves show a nearly identical evolution.}
\label{am10_compare}
\end{figure}

As indicated by these results, the standard box size of $4H\times8H\times8H$ is large enough for all of our simulations except for Am $\le$ 10.  It is computationally too expensive for us to run all of our simulations at the larger $8H\times16H\times8H$.  So, we elect to use the smaller size as our standard, and we compare the evolution of the stress between the $4H\times8H\times8H$ and $8H\times16H\times8H$ domains for Am = 10 to justify using a smaller domain for comparison between Am = 10 and other Am values.  Figure~\ref{am10_compare} shows the W$_{R\phi}$ evolution for these two domain sizes with Am = 10.  The use of a smaller domain size does not appear to make a difference for this value of Am.  Furthermore, the $B_y$ space-time plot for the larger domain looks very similar to the smaller domain.  Evidently, we can get away with using a smaller domain for Am = 10.  However, these ACFs suggest that caution be used when running ambipolar diffusion shearing box calculations.

One final thing to note from ACF($B$) is that as Am is decreased, the tilted centroid component appears to become more elongated (hence the need for larger domains) {\it and} less tilted with respect to the $y$ axis.  

To summarize this section, we find the turbulent stress level dependence on Am in vertically stratified simulations to be generally consistent with the results of
unstratified simulations \citep{bai11a}; $\alpha$ increases with Am, and for Am $\lesssim 1$, there is no turbulence.   We also find that as Am is decreased, larger domain sizes are needed in order to properly capture the turbulent structures represented by the ACF.  These runs serve as a baseline for interpreting the results from the next section, in which Am varies spatially and temporally based upon our model for surface layer ionization.

\subsection{Variable Am Calculations}
\label{variable_am}

We now turn our attention to the two calculations with spatially and temporally varying Am
(the ``Z" simulations in Table~\ref{tbl:sims}). As stated previously, these simulations adopt
more realistic non-constant Am values and directly address the questions such as ``how vigorous
is outer disk MRI-driven turbulence?" or ``what is the mass accretion rate in the outer disk?"
under the assumption that the outer disk is not threaded by any net vertical magnetic field.
We run them all at the largest domain size, $8H\times16H\times8H$, because there are
regions of Am~$\leq~1$ in these calculations.

\begin{figure}
\begin{center}
\includegraphics[width=0.5\textwidth,angle=0]{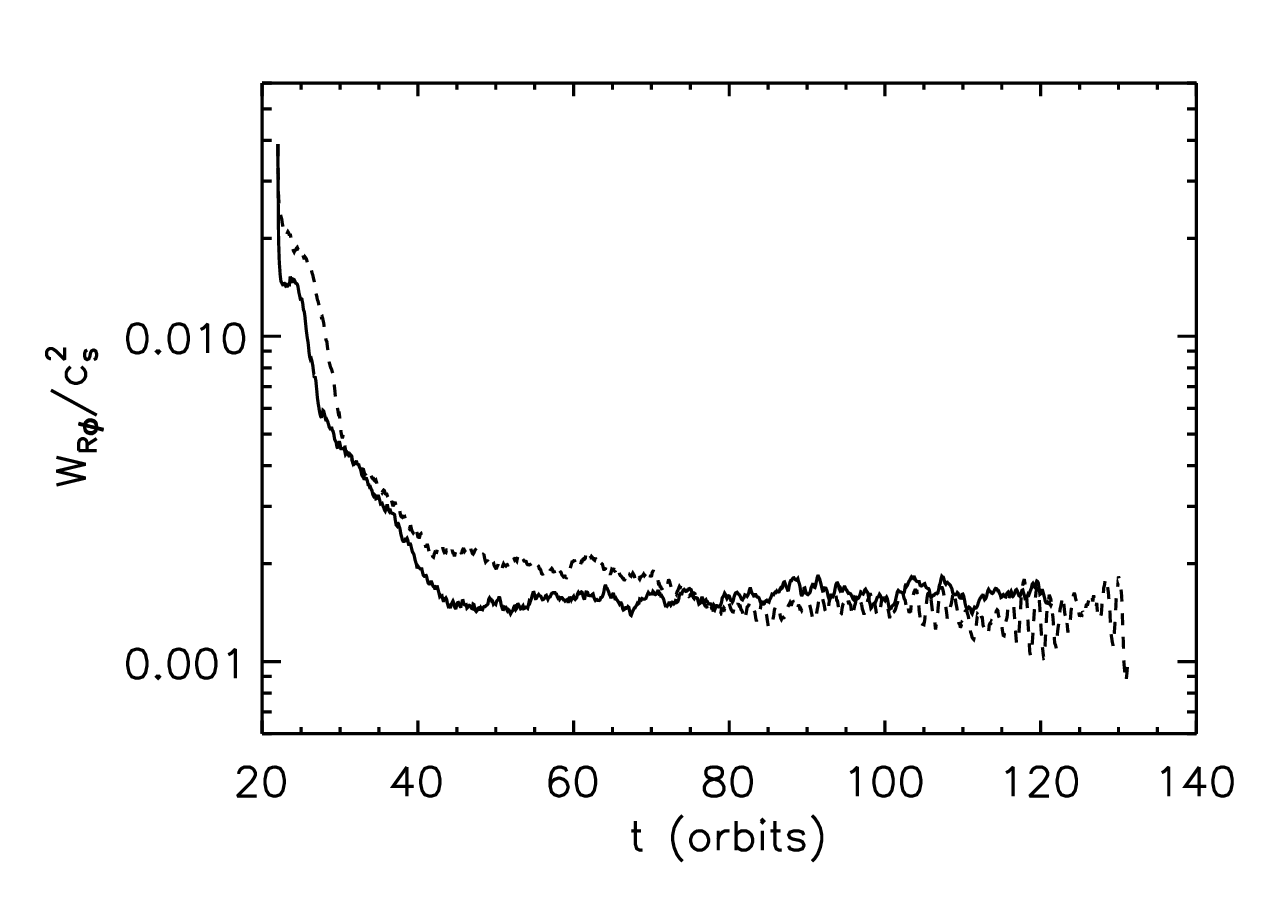}
\end{center}
\caption{Density-weighted volume average of the total (Maxwell and Reynolds) stress, normalized by the square of the sound speed, versus time for the Z30AU simulation (solid line) and the Z100AU simulation (dashed line).  The curves show a clear drop from the initial state of vigorous turbulence to levels near $W_{R\phi}/\cs^2 \sim 10^{-3}$.}
\label{z_hist}
\end{figure}

As before, we begin by examining the density-weighted stress normalized by the square of
the sound speed, as shown in Fig.~\ref{z_hist}.  It is clear from the figure that the stress levels
are lowered dramatically compared to the ideal MHD case.  Indeed, by calculating the
average of the curve from orbit 72 onward, we find that the $\alpha$ values are $\sim 10^{-3}$,
one order of magnitude below what is expected from observations \citep{hartmann98a}.

\begin{figure*}
\begin{center}
\includegraphics[width=0.75\textwidth,angle=0]{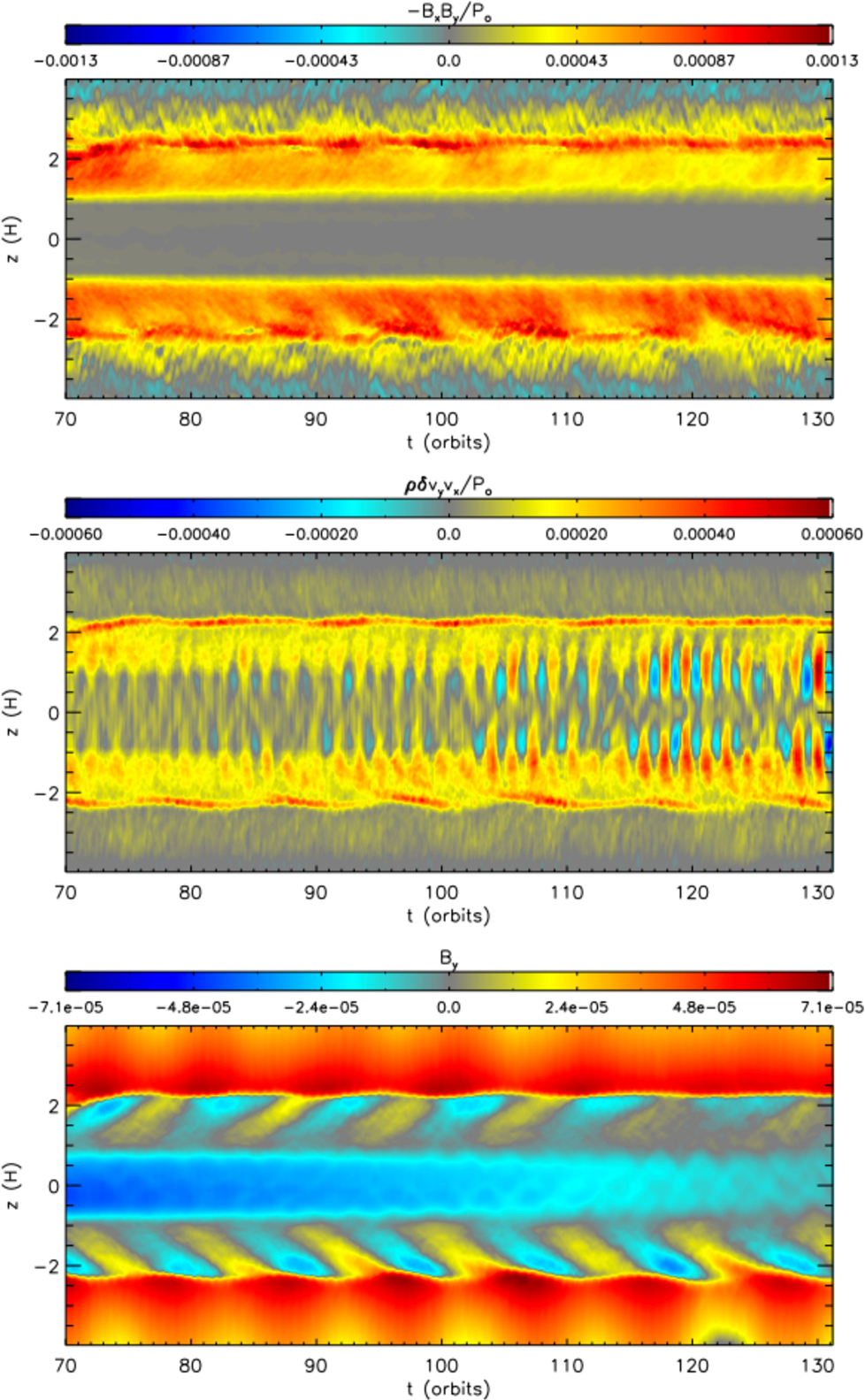}
\end{center}
\caption{Space-time plots of Maxwell stress (top), Reynolds stress (middle) and toroidal field, $B_y$ (bottom)
for run Z100AU, where each quantity has been averaged over all $x$ and $y$.  The stresses are normalized
by the initial peak gas pressure ($P_o = 5\times10^{-7}$), whereas $B_y$ is in code units.  There is a clear ``active" region between
1 and 2 scale heights above and below the mid-plane, where the Maxwell stress is non-negligible and the
toroidal field dynamo is active.  Within 1 scale height of the mid-plane, there is no turbulent activity, though there is non-zero
Reynolds stress.}\label{sttz_rau100}
\end{figure*}

While the turbulence initially decreases drastically, it has not completely died away.  Consider the space-time
diagrams in the $(t,z)$ plane for various horizontally averaged quantities of run Z100AU, as shown in
Fig.~\ref{sttz_rau100}.  From both the Maxwell and Reynolds stress plots, there is
a significant region in $z$ over which the MRI is indeed active.  The vertical structure is
consistent with what we would expect from our ionization profile; there is a significant ``ambipolar
dead zone" corresponding to where Am = 1 and the higher Am regions correspond to turbulent
activity.  We will calculate an actual mass
accretion rate at both 30 AU and 100 AU in Section~\ref{mass_acc}.

Despite there being very little Maxwell stress in the ambipolar dead zone, there is still Reynolds stress in this region.  This stress is likely produced by the active layers, similar to what is observed in simulations that include an Ohmic dead zone \citep{fleming03}.  It is possible, however, that as in run C1, the Reynolds stress here is simply left over from the turbulent state from which the run was restarted.  To test this notion, we restarted Z100AU at orbit 85 and set all perturbed velocity components to zero throughout the domain (the shear profile is of course maintained).  We find that the active MRI layers do indeed induce velocity fluctuations within the ambipolar dead zone, which leads to a positive Reynolds stress on average.

Finally, the horizontally averaged $B_y$ space-time plot shows interesting behavior.  Within the ambipolar dead zone, the toroidal field remains fixed in sign, though the magnitude appears to be decreasing.  Between 1 and 2 scale heights above and below the mid-plane, the usual MRI dynamo reappears, with a $B_y$ oscillation period of $\sim 10$ orbits, identical to ideal MHD.  Again, from the Maxwell stress plot, it is obvious that this region is turbulent.  For $|z| > 2H$, however, there is a strong toroidal field that remains stationary.  There are slight changes in magnitude as the toroidal field from the turbulent region propagates outwards.  

Examining the same diagrams for Z30AU reveals very similar behavior.  There is a thin layer of strong Maxwell and Reynolds stress located around $|z| \sim 2H$, along with a positive Reynolds stress induced in the ambipolar dead zone by the active layers.  In this calculation, however, the active region is much narrower in $z$ when compared to Z100AU.  This is because the FUV photons do not ionize the disk quite as deep as at $R = 100$ AU, and as is usual, the MRI appears to be inactive for $|z| \gtrsim 2H$.  Thus, the MRI-active region is forced to be contained within a smaller $z$ region.

In \cite{simon11a}, the authors found that the toroidal and radial field in a resistivity dominated mid-plane
region evolved in time. Occasionally, the toroidal field grew to
strong enough levels that the MRI became temporarily reactivated, after which the Ohmic
resistivity quenched the turbulence again.  This variability occurred on very long timescales of $\sim 100$ orbits.  In these simulations,
we do see that the toroidal field in the ambipolar dead zone changes in amplitude over time.
Is it possible, then, that the field could grow to large enough values to eventually re-activate
the MRI in that region?  Integrating our simulations out for many hundreds of
orbits is prohibitively expensive, and therefore, we will not be able to observe any such variability
in our stratified simulations.  Instead, we have run an {\it unstratified} shearing box of size
$8H\times16H\times H$ with Am = 1 and uniform toroidal field of strength $\beta = 10$, chosen
to determine if a strong toroidal field MRI will be active with Am = 1.  We start the simulation
with a relatively large amplitude perturbation to the density and velocities, such that the initial
perturbations should already be reasonably nonlinear.  As a control, we also ran this identical
setup with Am = $10^5$. With Am = 1, we observed decay of the initial perturbations and no
development of any MRI turbulence, whereas with Am = $10^5$, the MRI becomes active and
generates sustained turbulence for many orbits. Thus, even in the presence of a strong field,
Am = 1 is sufficient to quench MRI-driven turbulence.  We therefore do not expect the
behavior observed in \cite{simon11a} to occur in these simulations.

\begin{figure*}
\begin{minipage}{8cm}
\begin{center}
\includegraphics[width=1\textwidth,angle=0]{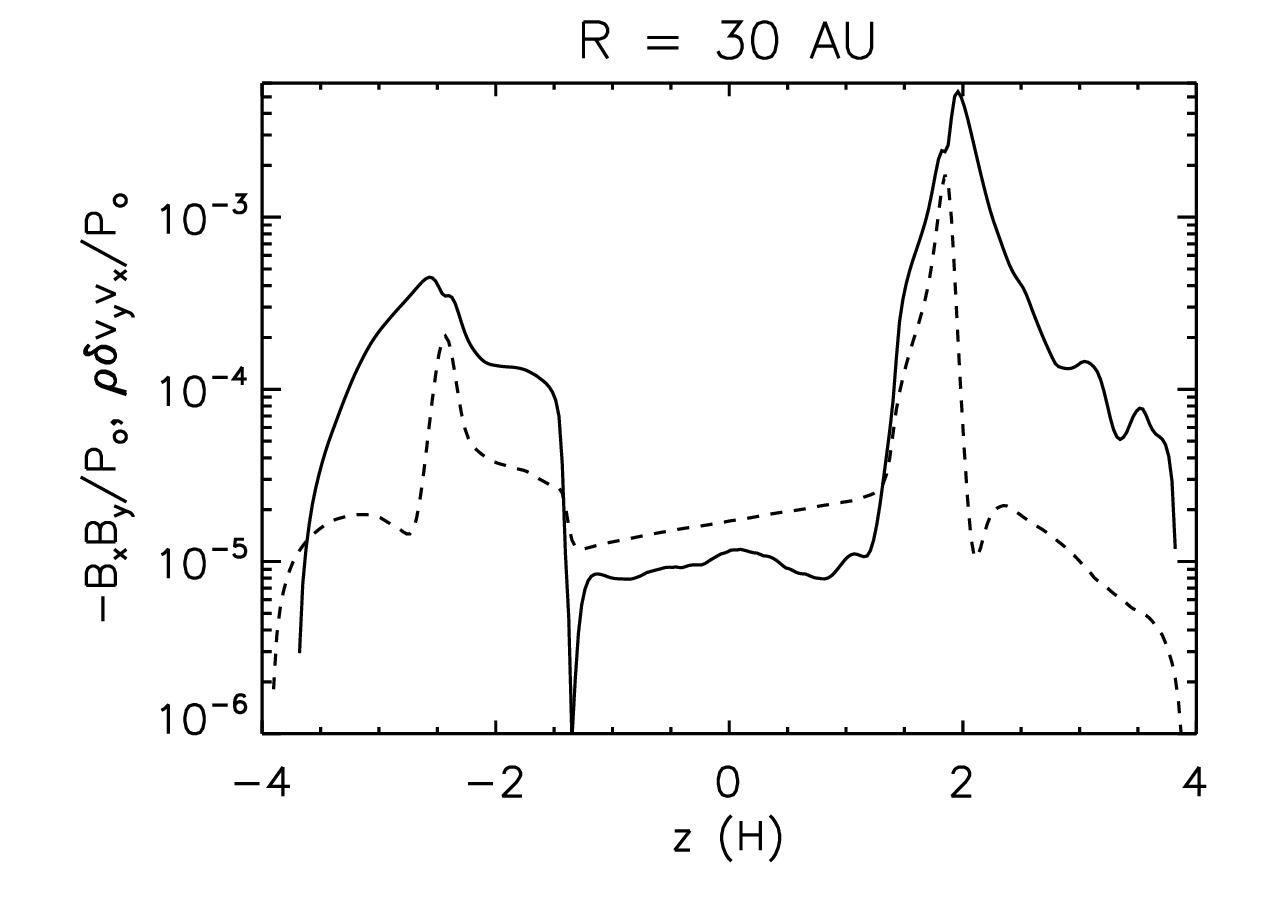}
\end{center}
\end{minipage}
\begin{minipage}{8cm}
\begin{center}
\hspace{-3.8cm}
\includegraphics[width=1\textwidth,angle=0]{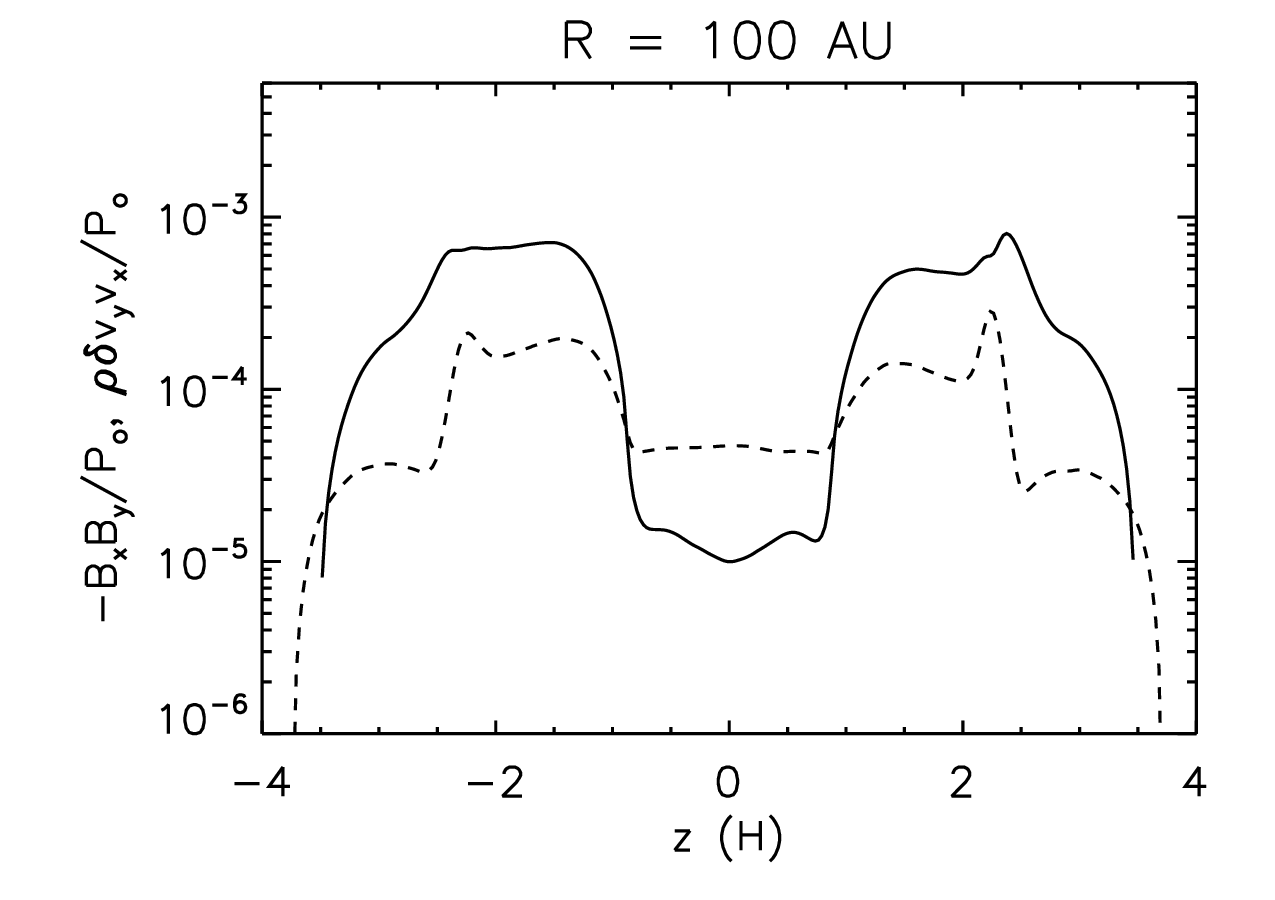}
\end{center}
\end{minipage}
\begin{minipage}{8cm}
\begin{center}
\includegraphics[width=1\textwidth,angle=0]{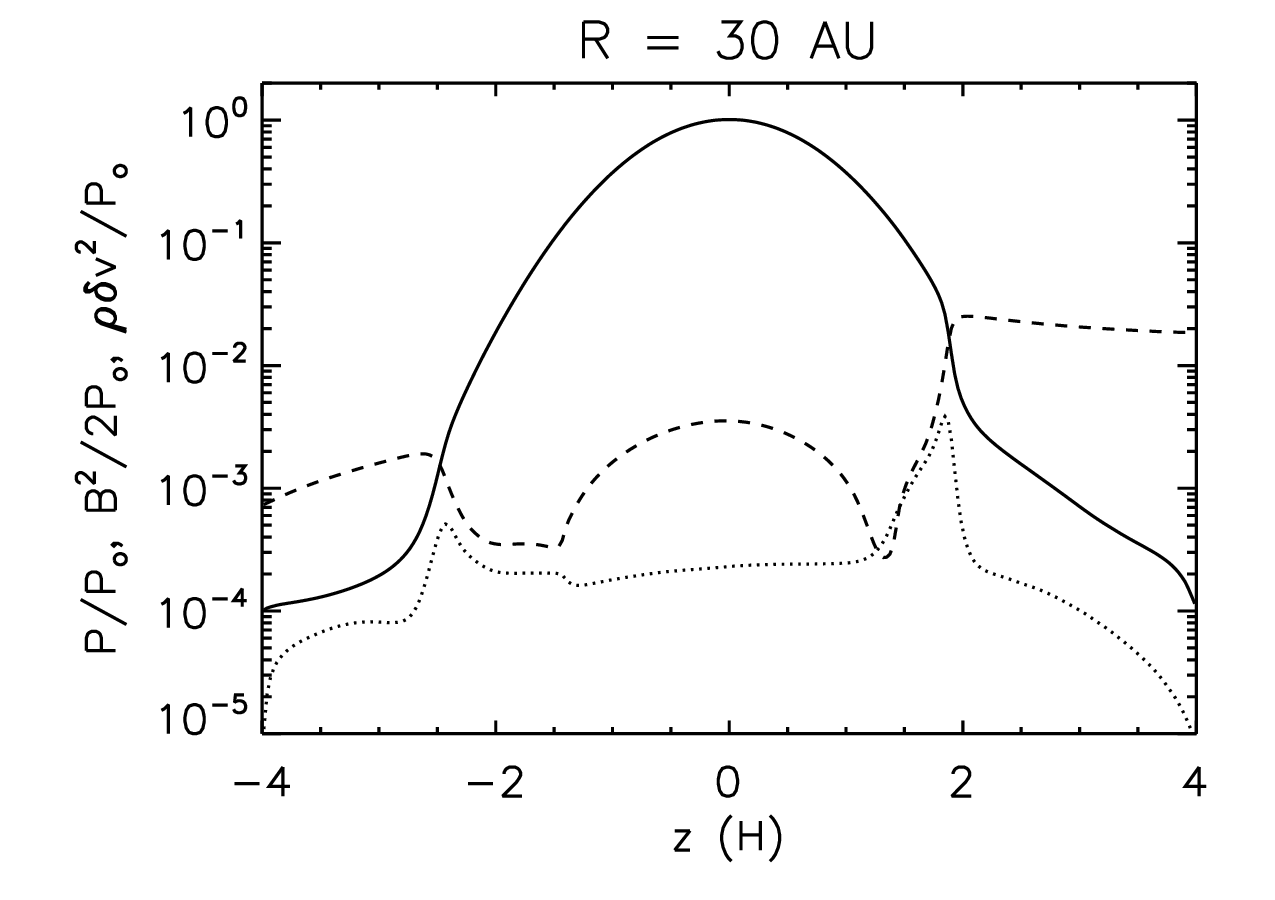}
\end{center}
\end{minipage}
\begin{minipage}{8cm}
\begin{center}
\includegraphics[width=1\textwidth,angle=0]{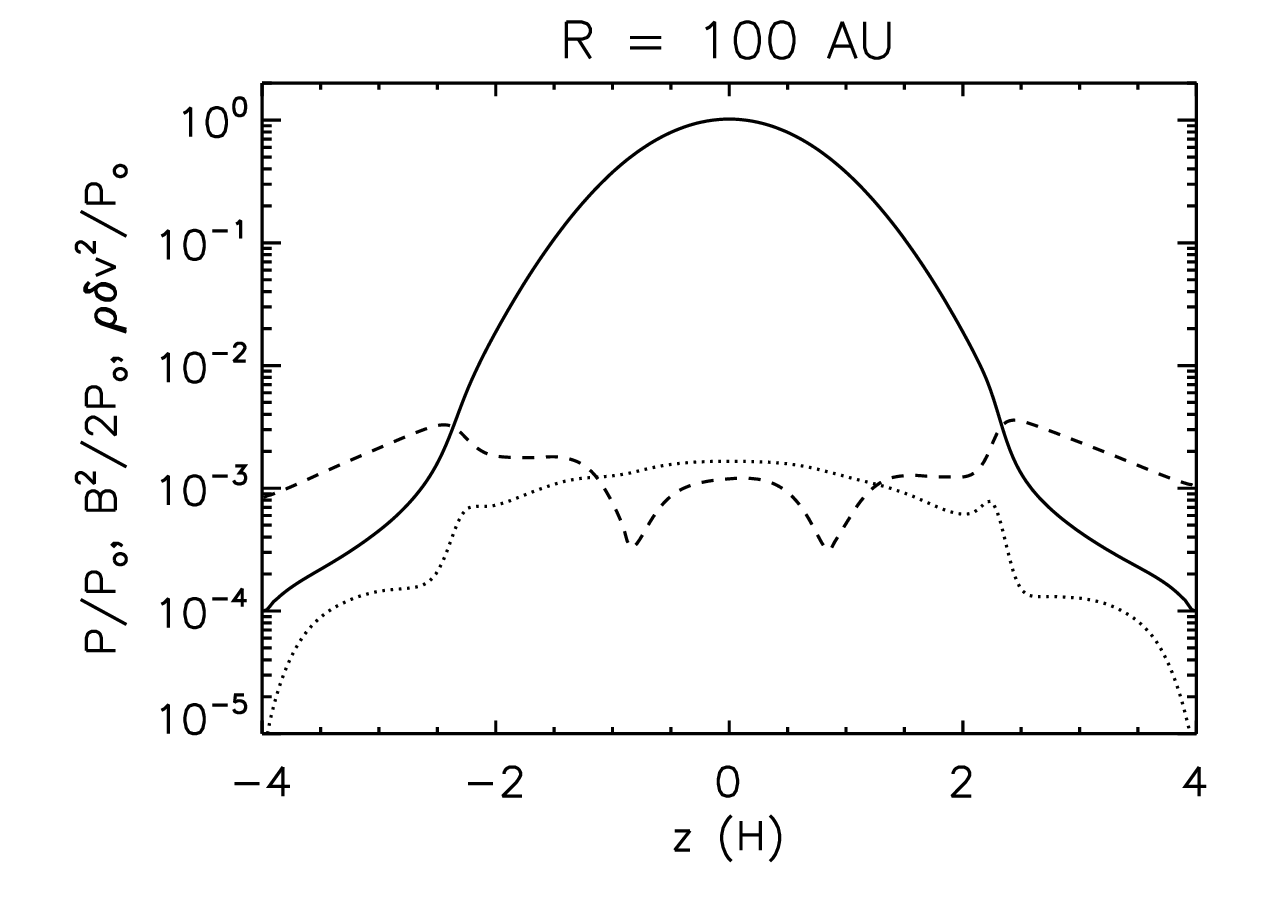}
\end{center}
\end{minipage}
\caption{Time- and horizontally-averaged stresses (upper row) and energies (bottom row) vs. $z$ in scale heights for the run at 30 AU (left column) and 100 AU (right column).  The time average is done from orbit 72 onwards, and the horizontal average is done over all $x$ and $y$.   In the stress plots, the solid line is the Maxwell stress, and the dashed line is the Reynolds stress.  In the energy plots, the solid line is the gas pressure, the dashed line is the magnetic energy, and the dotted line is the kinetic energy.  All quantities are normalized by the initial peak gas pressure ($P_o = 5\times10^{-7}$).}
\label{vert_profile}
\end{figure*}

The vertical structure is also depicted via time- and horizontally-averaged quantities as a function of $z$, as is shown in Fig.~\ref{vert_profile}.  From the top two panels, it is obvious that there is a double-peak structure to the stress; no doubt a result of the ambipolar dead zone.  Within this dead zone, the Reynolds stress dominates over the Maxwell stress. From the space-time diagrams above, any nonzero Maxwell stress within this region likely results from large scale correlations in the magnetic field rather than any sort of turbulence.  Note that there are regions where the stress can go negative, and since the vertical axis is logarithmic, the curves are simply cut off where the values drop below zero.

The bottom two panels show the various energies (i.e., thermal, kinetic, and magnetic) as a function of height. The thermal pressure dominates over the vast majority of the disk's vertical structure.  However, for $|z| \gtrsim 2.5H$, the magnetic energy dominates over all other energies.  We note in particular the very strong magnetic dominance in the upper $z$ regions of Z30AU (lower left panel).  This magnetic field is stationary in time according to the space-time diagram for this run, akin to the lower panel of Fig.~\ref{sttz_rau100}.  

We attribute some of this behavior to the gas density floor.  Indeed, looking at the vertical pressure profile (which follows the gas density), the pressure is prevented from going below $10^{-4}$ of its initial peak value.  The resulting gradient in the gas pressure has an effect on the buoyant properties of the field.  We address this issue further
in the next section.

\subsection{The Effect of the Density Floor}
\label{density_floor}

To test the effect of lowering the density floor, we restarted simulations C10 and Z30AU at orbits 120 and 56 respectively and lowered the density floor to $3\times10^{-5}$, roughly 3 times smaller than the original density floor.  While this new value for the floor may still be higher than what the density would naturally be, it becomes extremely expensive to run lower density floor calculations even for a short integration.  Thus, our main goal in running these calculations is to observe the immediate effect of lowering the floor on properties such as the density-weighted stress and the buoyancy of the magnetic field in the upper disk regions.  Is the evolution of the system altered drastically?  What, if any, changes occur?

Examining the space-time diagrams for these restarted runs indicate that the lowered density floors lead to enhanced buoyancy of the magnetic fields.  For example, consider the lower left panel of Fig.~\ref{vert_profile}.  Once the floor is lowered, the magnetic energy for $z \lesssim -2H$ immediately drops to lower values as field rises away from the mid-plane.\footnote{The field for $z \gtrsim 2H$ does not appear to change significantly over the time that we integrated this simulation.  However, we did not  evolve this simulation for very long, and we speculate that a longer evolution would show a change in the magnetic field strength in this region.} The same behavior is observed in run C10 for both $z \lesssim -2H$ and $z \gtrsim 2H$.  Thus, the presence of a strong magnetic field for $|z| > 2H$, as is shown in the lower panel of Fig.~\ref{sttz_rau100}, is artificial.  That being said, based upon the results of ideal MHD calculations \citep{simon11a} in which the density is on average larger than the floor value at all locations, we still expect the field to be superthermal in this region; it will just be weaker.

Does this feature affect our main results so far?  An examination of the stress evolution shows that the decrease in the density floor does lead to a decrease in the volume-averaged stresses.  Quantitatively speaking, in Z30AU, the magnetic energy averaged over $z \le -2H$ drops by a factor of $\sim 4$ in going from orbit 56 to 78.5.  During this same time, the volume-averaged stresses decrease by a factor of $\sim 1.5$, and the stress does not appear to be leveling off.  Running the simulation out further is computationally prohibitive given the small time step incurred by the lower density floor.  So, there {\it is} an effect due to the density floor.  Since we have not fully quantified this effect, it should be borne in mind when we calculate mass accretion rates in Section~\ref{mass_acc}.   We address the density floor issue again in Section~\ref{turb_linewidth}.

\section{Implications for Protoplanetary Disk Structure and Evolution}
\label{implications}

\subsection{Mass Accretion Rate}
\label{mass_acc}

One of the most important properties of disk evolution is the mass accretion
rate due to angular momentum transport. We can calculate this quantity, $\dot{M}$, for Z30AU
and Z100AU by utilizing equation (40) in \cite{balbus98}, which assumes accretion is in steady-state.
We take this equation in the limit that $R \gg R_o$, as appropriate for the shearing box approximation:
\begin{equation}
\label{mdot}
\dot{M} = \frac{2\pi\Sigma}{\Omega}\overline{W_{R\phi}} \approx 8.5\times10^{-6}\alpha R_{\rm AU}^{-1/2}\ M_{\sun}/{\rm yr},
\end{equation}

\noindent
where the expression on the right comes from applying this formula to the minimum mass solar nebula model, and $\alpha$ is defined as in equation~(\ref{alpha}).
The definition of $W_{R\phi}$ in \cite{balbus98} is the same as ours in the sense that it is a {\it density weighted}
stress. From this equation, we calculate that $\dot{M} \approx 2.5\times10^{-9} M_{\sun}/$yr at 30 AU and  $\dot{M} \approx 1.2\times10^{-9} M_{\sun}/$yr at 100 AU.
Of course, due to the effect from the high density floor (see discussion in Section~\ref{density_floor}), the actual accretion rates are likely to be lower, and these values serve as an upper limit.  

These accretion rates are at least one order of magnitude too small (likely even smaller, again due to the density floor) to account for the
observed accretion rates in Classical T-Tauri systems \cite[e.g.,][]{hartmann98a}.  

We can also compare our results to semi-analytical predictions made by \cite{bai11b} based upon the results of
unstratified simulations with a net vertical magnetic flux \citep{bai11a}.   This comparison will allow us to gauge the potential
importance of having a net vertical flux in regions of strong ambipolar diffusion.
By first extracting the vertical profiles of $\rho$ and Am, and then by assuming constant magnetic field strength
across the MRI active region of the disk, we can estimate the accretion rate for any given
field strength using equation (28) of \cite{bai11b}.  From this approach, $\dot{M} = 9.8\times10^{-9}M_{\sun}/$yr and $3.5\times10^{-8}M_{\sun}/$yr at 30 AU and 100 AU, respectively, roughly one order of magnitude larger than our calculated rates and in general
agreement with observations.

These estimates suggest that it is very likely that a net vertical field is required to
attain the necessary turbulence levels, if the MRI is indeed the dominant mechanism by
which angular momentum is transported. We will carry out actual shearing box simulations with vertical stratification and ambipolar
diffusion in the presence of a net vertical magnetic field in Paper II.

\subsection{Turbulent Linewidth}
\label{turb_linewidth}

Another property of disks of recent interest has been the density-weighted distribution of turbulent
velocities as a function of disk radius and height above the mid-plane.  This was first calculated by
\cite{simon11b} for local MRI simulations without ambipolar diffusion, but including the effects of Ohmic
resistivity.  Another study, \cite{forgan12}, carried out a similar analysis for global calculations of
self-gravity driven turbulence.  These distributions are a first order approach to making a connection with
observational constraints of turbulent line broadening in the sub-mm, such as those in
\cite{hughes11}.   In particular, the density-weighted velocity distribution represents the probability of observing a
line with a particular turbulent velocity broadening.   

\begin{figure}
\begin{minipage}{8cm}
\begin{center}
\includegraphics[width=1.1\textwidth,angle=0]{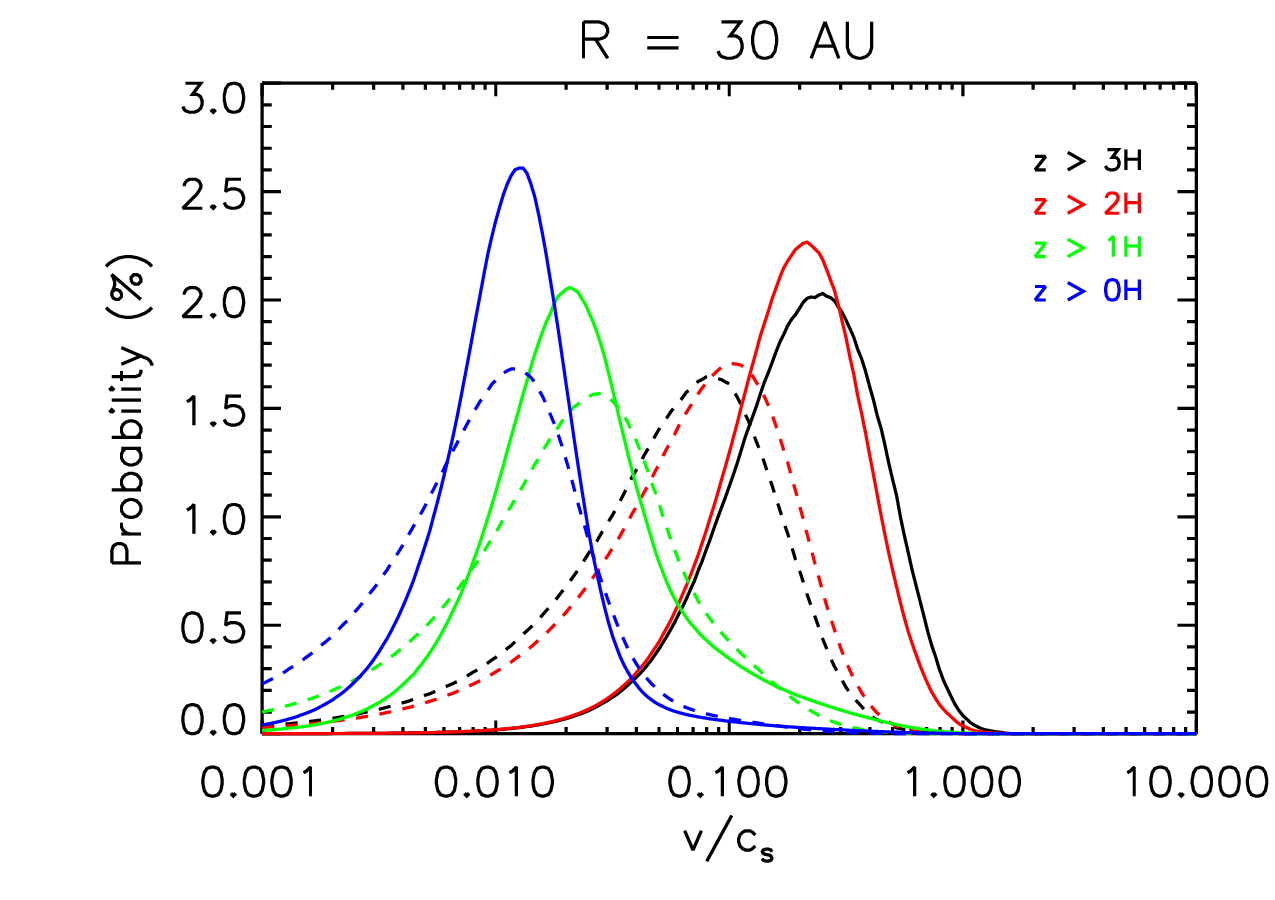}
\end{center}
\end{minipage}
\begin{minipage}{8cm}
\begin{center}
\includegraphics[width=1.1\textwidth,angle=0]{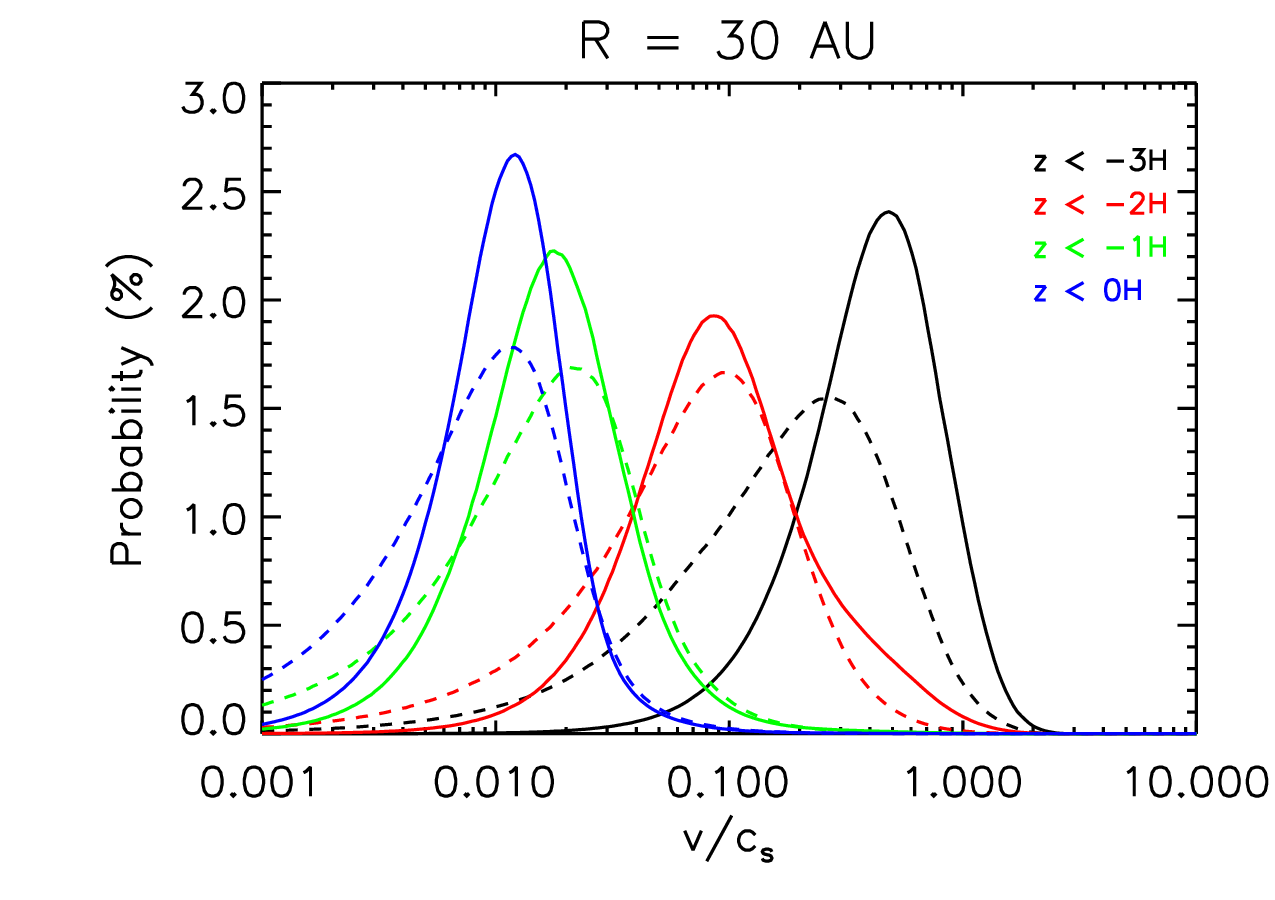}
\end{center}
\end{minipage}
\begin{minipage}{8cm}
\begin{center}
\includegraphics[width=1.1\textwidth,angle=0]{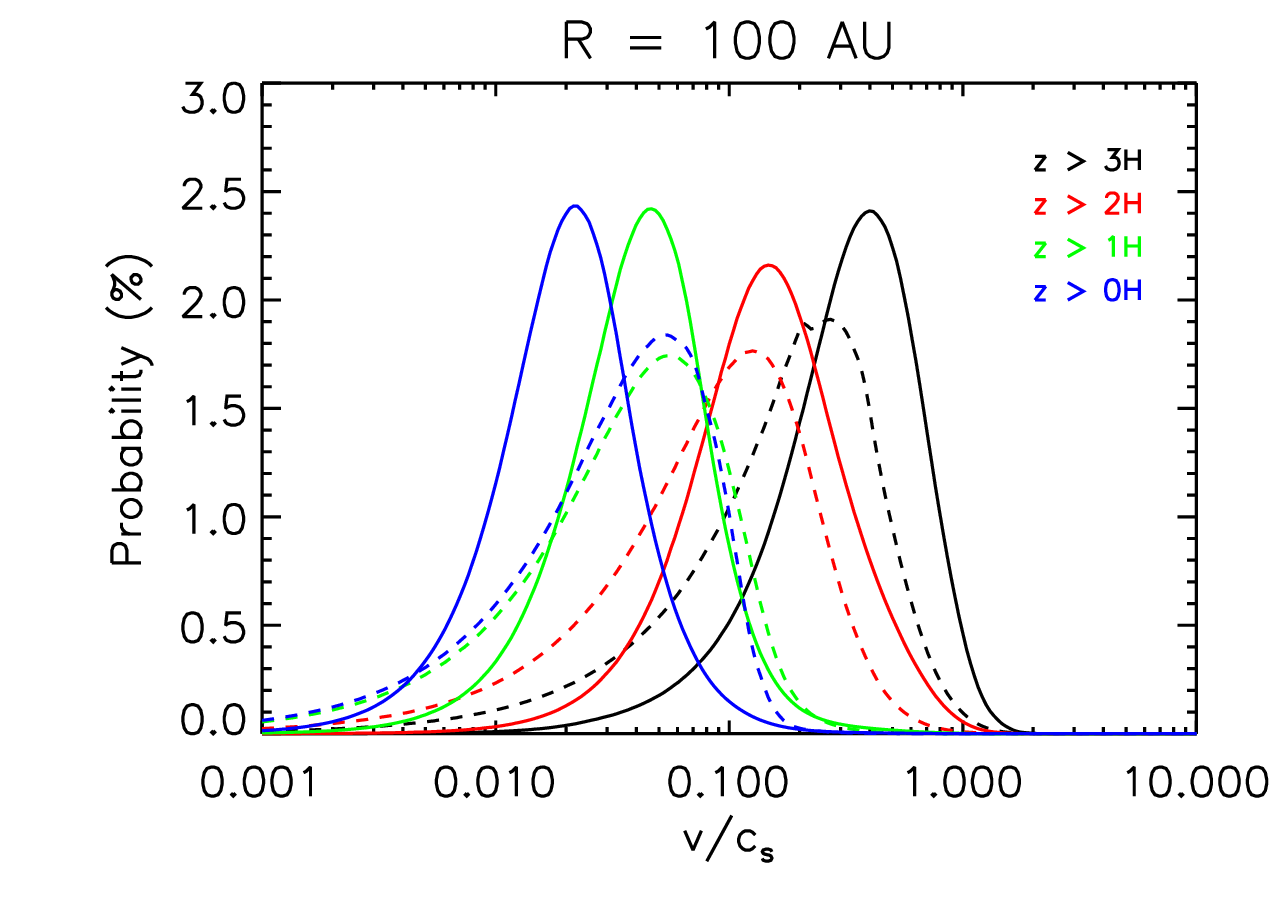}
\end{center}
\end{minipage}
\caption{Density-weighted turbulent velocity distributions for Z30AU (top and middle) and Z100AU (bottom).  The top and bottom panels correspond only to velocities at $z > 0$, whereas the middle panel corresponds to $z < 0$.  Each color corresponds to different depths over which the distribution is calculated, as labeled. The dashed lines are the vertical turbulent velocity $|v_z|/\cs$, and the solid lines are the azimuthally averaged disk planar velocity $|v_h|/\cs$.  Relatively small velocities exist towards the mid-plane, as turbulence is very weak there.  However, far from the mid-plane, the velocity consistently peaks at $v/\cs \sim 0.2-0.4$, with part of the distribution going into the supersonic regime.}
\label{vturb}
\end{figure}

Here, we carry out an identical analysis to that done in \cite{simon11b} for both of our variable Am
calculations (see that paper for the exact details of how to calculate the velocity distribution). 
Figure~\ref{vturb} shows this velocity distribution for Z30AU (top and middle) and Z100AU (bottom).   As was
observed in \cite{simon11b} for their calculations with a strong Ohmic dead zone (see their Fig. 4, top
panel), we also observe a strong gradient in peak velocity as one probes
deeper towards the mid-plane.  Indeed, the mid-plane velocity distribution peaks around $v/\cs \sim 0.01$,
just as in the Ohmic case.  Furthermore, as one probes the surface layers of the disk $|z| > 3H$, the peak of
the distribution occurs around $v/\cs \sim 0.2-0.4$.  There is also a non-negligible supersonic tail; in Z100AU,
this component comprises $\sim 1\%$ of the horizontal and vertical distributions at $z > 3H$.  In Z30AU, 
this component comprises $\sim 2\%$ of the vertical distribution and $\sim 7\%$ of the horizontal distribution at $z < -3H$. 

In the top panel of the figure, the red and black curves nearly lie on top of each other.  This is likely an artificial effect resulting from the very dominant magnetic field that is stationary for $z > 2H$ in Z30AU (which itself results from the relatively large density floor applied in this calculation).  The magnetic field is not nearly as dominant for $z < -2H$ (the other side of the mid-plane), and so the middle panel of Fig.~\ref{vturb} shows the velocity distribution calculated from this side.  This distribution looks much more similar to the other distributions.  

To further test the effect of the density floor on our velocity distributions, we have rerun Z30AU with the density floor lowered to $3\times10^{-5}$ (as discussed in Section~\ref{density_floor}).  We calculated the velocity distribution for this region during two separate periods, each averaged over 8 orbits.  We do not see any significant difference between these velocity distributions and that shown in the middle panel of Fig.~\ref{vturb}. 

Finally, the rough agreement between turbulent velocity distributions for the vertical velocity and the ``in-plane" velocities suggest that turbulent motions will be isotropic, consistent with previous results \citep{simon11b}.  We point out that velocity distribution for $z > 0$ in Z100AU suggests that the flow is slightly anisotropic.  We are not  entirely sure why this is the case.  However, when we restarted this run and set the turbulent velocity to zero (as described in Section~\ref{variable_am}), the resulting velocity distribution was again nearly isotropic.  Since this isotropy is present in all of the other cases, it seems more likely that the distribution for $z > 0$ in Z100AU is a peculiar case, perhaps resulting from the exact nature of the turbulent state from which this run was initiated.

\section{Summary and Conclusions}
\label{conclusions}

We have run local shearing box simulations of MRI-driven turbulence in the
presence of ambipolar diffusion and in the absence of a net vertical magnetic
field. These simulations were designed to address two primary questions: 

\begin{itemize}

\item How does MRI-driven turbulence behave in the presence of both ambipolar diffusion and vertical gravity?

\item What are the implications for turbulence in the outer regions of protoplanetary disks where ambipolar diffusion is dominant? 

\end{itemize}

With the ambipolar Elsasser number, Am, (see equation~\ref{am1}) remaining constant, we addressed
the first question. We found that the density-weighted stress decreases with increasing ambipolar diffusion
and decays to negligible values for ${\rm Am} < 1$.  Another parameter that controls the stress levels,
however, is the amplitude of the toroidal field strength, which oscillates in time due to the dynamo
\citep{simon11a}. The stronger the amplitude of oscillation (but not so strong that the MRI is suppressed
in the presence of ambipolar diffusion), the larger the turbulent stresses. The average
strength of this varying field combined with ambipolar diffusion affect the MRI in such a way that the
turbulent stress does not necessarily increase monotonically with decreasing diffusion.
All our results are consistent with unstratified numerical simulations of \cite{bai11a}, although the
subtleties of the ambipolar MRI dynamo due to the addition of vertical gravity does not guarantee
a one-to-one correspondence between the level of diffusion and that of the turbulent stress.

Additional noticeable effects emerged from these constant Am calculations.  First, as ambipolar diffusion is
increased, the dynamo oscillation period becomes longer.  Above Am $\sim 100$, the oscillation period
approaches the ideal MHD limit of 10 orbits. This result opens up more questions than it answers, as we do
not yet have an understanding of the dynamo in the ideal MHD limit, let alone including a diffusion term. 
However, it may be insightful to apply an ambipolar diffusion term to current, simplified models of the MRI
dynamo to help in further understanding the physics of the dynamo.   Second, the typical turbulent fluctuations become larger in scale (i.e., have a larger correlation length) and become more aligned with the azimuthal direction as ambipolar diffusion is increased. 
This last point has important implications for local simulations that include ambipolar diffusion. Increased
diffusion evidently favors larger scale fluctuations, and for simulations with strong diffusion, larger
shearing boxes are required.  A box size of at least $8H\times16H\times8H$ is required to properly capture,
the MRI turbulence with Am $\lesssim 10$. This also motivates further studies of ambipolar diffusion and the MRI 
in {\it global simulations}, where one is not limited by scales of order $H$.

To answer the second of our motivating questions, we ran additional simulations that included a physically
motivated model for the ionization structure (and hence Am profile) of the protoplanetary disk. These runs
include the effect of a strong FUV ionization layer based upon the work of \cite{perez-becker11b}, where
a large ionization fraction $f\sim10^{-5}$ can be achieved in a very thin layer above and below the disk
mid-plane, while the rest of the disk is assumed to have $Am=1$. Although this ionization model still bears uncertainties, it provides the
essential physical ingredients that allow us to explore the gas dynamics in the outer regions of protoplanetary disks
in a realistic manner, again assuming zero vertical magnetic flux.

We find that despite this strong FUV ionization, the mass accretion rate is of order $10^{-9} M_{\sun}$/yr,
too small to account for observed accretion rates measured in T~Tauri
stars \citep{gullbring98,hartmann98a}.   In fact, this estimate should be treated as an upper limit due to the increase in stress from the relatively large density floor employed. This small accretion rate and the presence of the ambipolar dead zone is reminiscent of models for the
Ohmic dead zone \citep{gammie96}, many of which also yield low accretion rates. The problem
posed by an ambipolar dead zone is, however, more serious. Close to the star, the viscous time
scale $R^2/\nu$ can be short compared to the disk lifetime given even a weak residual stress. If
an Ohmic dead zone can be supplied with gas from further out, it is then possible to imagine gas
accumulating there until some additional instability allows gas to flow through on to the star
\cite[e.g.,][]{armitage01,zhu10,martin12}. In the ambipolar dead zone, conversely, low levels of
stress imply that the main mass reservoir is permanently inactive. Since this appears to contradict observations,
our simulations are either missing some important aspect of the physics, or our basic understanding of disks
is incorrect.

There are several possibilities. First, we could be missing some effects that strongly influence the MRI. Including the Hall effect
may substantially enhance the strength of MRI turbulence, as indicated by previous works (see the Introduction). Except
for the simulations of \cite{sano02a} and \cite{sano02b}, no one has carried out non-linear MRI simulations of the Hall effect.  
Somewhat discouragingly, though, the results of the Sano studies suggest that the Hall effect does not have a strong influence
when Ohmic resistivity is dominant.  The same could conceivably be true for ambipolar diffusion dominated regions of the disk
as this diffusion acts similar to Ohmic resistivity; it damps out the turbulence.  

Alternatively, it is possible that angular momentum transport in the outer disk is dominated by
an entirely different physical mechanism, though the known candidates are not expected to be
efficient in this region unless the disk mass is high enough that self-gravity is significant
\citep{armitage11}. It could also be the case that the surface density drops off less rapidly
than that assumed in our model. If this is the case, then the induced Reynolds stress from the
active layers {\it may} be able to transport more angular momentum outwards, thus increasing
the accretion rate.  However, a larger column in the outer disk leads to a smaller active region
and thus a smaller induced Reynolds stress.  Finally, one cannot exclude the possibility that the
disk on 30-100~AU scales is genuinely inviscid, with observed accretion coming from a
larger-than-expected reservoir of gas closer to the star.

However, by far the most promising possibility for explaining these low accretion rates is the exclusion of a net vertical magnetic field.
Not only is the inclusion of a net vertical field the most optimal magnetic geometry for the MRI with ambipolar diffusion \citep{bai11a}, but it is very
likely that the disk would be penetrated by at least some amount of vertical field.
This vertical field allows the MRI to operate at small values of Am and permits relatively strong turbulence with $\alpha\sim0.01$ at the
mid-plane region of the outer disk \citep{bai11b,bai11c}.  It should also make the accretion in the
 FUV ionized layer much more efficient since the vertical field will be relatively strong here due to the drop in gas pressure away from the mid-plane. Indeed, an estimate of the accretion rate based on the models of  \cite{perez-becker11b} and \cite{bai11a} and
the simulations of \cite{bai11a} (which include the effects of a net vertical magnetic flux) returns a much more optimistic $\dot{M} \sim 10^{-8} M_{\sun}/$yr.  These considerations strongly motivate our companion paper, where we will study in detail the effect of net vertical fields on the accretion process in the limit of
strong ambipolar diffusion. 

The outer regions of protoplanetary disks can be resolved at sub-mm wavelengths, and with
this in mind we  have calculated the probability distribution for turbulent velocities as a function
of height within the disk. We find that, although the ambipolar dead zone severely restricts the
accretion rate, turbulence in the active surface layers remains strong. We obtain peak values of
$\sim 0.4$ of the isothermal sound speed. These results are similar to those found by
\cite{simon11b} in calculations of Ohmic dead zones at smaller radii, and are consistent with
observations of HD 162396 made by \cite{hughes11}. We predict that the turbulent velocity
(and hence line width) ought to be a strong function of height at radii where an ambipolar dead
zone is present, which may be testable given observations of multiple molecular tracers that
probe different depths within the disk.

The most promising avenue for observational progress lies in ALMA measurements of the spatial
structure and velocity field of protoplanetary disks on the same (large) scales as those considered
theoretically in this paper. As we have noted, improved measurements of turbulent line broadening
\citep{hughes11} at different depths within the disk can potentially provide direct constraints of
theoretical models. Such observations appear to be technically feasible (Hughes, private
communication). Our results, however, also motivate consideration of disk evolution scenarios
that are substantially different from those usually adopted in the interpretation of observational data.
It is commonly assumed, for example, that the outer edges of protoplanetary disks expand
significantly as the disk evolves viscously. If this is true, then measurements of the disk surface
density profile (when fit, for example, by similarity solutions) constrain the radial variation of the
angular momentum transport. Our results, on the other hand, suggest that some disks (those with
negligible net vertical fields) may not evolve viscously at all on large scales. It may therefore be
useful to consider models in which qualitatively different pathways of disk evolution are driven by
variations in the initial magnetic flux, rather than by differences in the initial mass and angular
momentum content of disk gas. Since strong vertical fields can also lead to angular momentum
loss via disk winds \cite[e.g.,][]{salmeron11,lesur12}, one possible scenario is that strong vertical fields lead
to wind-dominated disks, weaker fields to stimulated MRI-driven evolution, and no vertical field to
effectively inviscid disks.   

\acknowledgments

We thank Sean O'Neill and John Hawley for useful discussions and
suggestions regarding this work.  We also thank an anonymous referee for useful comments on an earlier draft of this paper. JBS, PJA, and KB acknowledge support from NASA through grants NNX09AB90G and NNX11AE12G and from
the National Science Foundation through grant AST-0807471.  KB also acknowledges funding support from Tech-X Corp., Boulder, CO.
 XB and JMS acknowledge support from the National Science Foundation through grant AST-0908269.  XB also acknowledges support for program number HST-HF-51301.01-A provided by NASA through a Hubble Fellowship grant from the Space
Telescope Science Institute, which is operated by the Association of Universities for Research in Astronomy, Incorporated, under NASA contract
NAS5-26555. This research was supported by an allocation of advanced computing resources provided by the National Science Foundation. The computations were performed on Kraken and Nautilus at the National Institute for Computational Sciences through XSEDE grant TG-AST090106.
This work also utilized the Janus supercomputer, which is supported by the National Science Foundation (award number CNS-0821794) and the University of Colorado Boulder. The Janus supercomputer is a joint effort of the University of Colorado Boulder, the University of Colorado Denver, and the National Center for Atmospheric Research.

\appendix

\section{Super Time-Stepping}

In this work, we have employed the super time-stepping (STS) technique of
\cite{alexiades96} to allow for an accelerated integration while including the effects of strong ambipolar diffusion.
Following \cite{osullivan06} and \cite{osullivan07}, the STS method divides a compound timestep $\Delta t_{\rm STS}$ 
into $N$ {\em unequal} substeps $\Delta \tau_j$ ($j=1,...,N$) with
\begin{equation}
\Delta t_{\rm STS}=\sum_{j=1}^{N}\Delta\tau_j\ .
\end{equation}
By choosing $\Delta \tau_j$ judiciously, stability can be maintained even when the 
averaged timestep $\Delta t_{\rm STS}/N$ is much larger than the normal
stable diffusion timestep $\Delta t_{\rm diff}$. The optimized
lengths for the substeps were found to be \citep{alexiades96,osullivan06,osullivan07}
\begin{equation}
\Delta\tau_j=\Delta t_{\rm diff}\bigg[(\nu-1)
\cos\bigg(\frac{2j-1}{N}\frac{\pi}{2}\bigg)+\nu+1\bigg]^{-1}\ ,
\end{equation}
where $0<\nu<1$ is a free parameter. The sum of the substeps gives
\begin{equation}
\Delta t_{\rm STS}=\Delta t_{\rm diff}\frac{N}{2\sqrt{\nu}}\bigg[
\frac{(1+\sqrt{\nu})^{2N}-(1-\sqrt{\nu})^{2N}}
{(1+\sqrt{\nu})^{2N}+(1-\sqrt{\nu})^{2N}}\bigg]
\equiv G(N,\nu)\Delta t_{\rm diff} .
\end{equation}
We note that as $\nu\rightarrow0$, 
$\Delta t_{\rm STS}\rightarrow N^2\Delta t_{\rm diff}$ so that the STS approach is
asymptotically $N$ times faster than the standard explicit approach. However,
the value of $\nu$ needs to be properly chosen to achieve the optimal balance
between performance  and accuracy.
In general, the STS method provides better accuracy as $N$ decreases
and $\nu$ increases, whereas large $N$ and small $\nu$ lead to higher efficiency.
Here, we choose $\nu=1/4N^2$ with a limit of $N\leq12$. At $N=12$, one
achieves an acceleration factor of about $9$. It is also found that further
increasing $N$ would not significantly increase the efficiency without sacrificing
accuracy (Stone, private communication based on a Princeton Junior Project done by Sara Wellons).

In our calculations, we first compute the ideal MHD time step $\Delta t_{\rm MHD}$
and the diffusion timestep $\Delta t_{\rm diff}$. The number of super time steps $N$ can be
found from the condition $G[N-1,1/4(N-1)^2]<\Delta t_{\rm MHD}/\Delta t_{\rm diff}\leq G(N, 1/4N^2)$.  
If $N\leq12$, then we modify $\Delta t_{\rm diff}$ so that $\Delta
t_{\rm MHD}\equiv G(N,1/4N^2) \Delta t_{\rm diff}$. Otherwise, we fix $N = 12$, and set
$\Delta t_{\rm MHD}=\Delta t_{\rm diff}G[12, 1/(4\times12^2)]$. In this
way, we always have $\Delta t_{\rm MHD}=\Delta t_{\rm STS}$.

 As we use an
operator-split algorithm for magnetic diffusion, we integrate $N$ STS substeps of
the ambipolar diffusion term with $\Delta \tau_j$ before evolving one MHD time step with
$\Delta t_{\rm STS}$. With STS, we have repeated the test problems (i.e., linear
wave damping test and C-type shock test) performed in \citet{bai11a} and found
essentially the same results for $N_{\rm STS}$ up to 10. Moreover, we repeated
the unstratified MRI simulations with $Am=1$ (runs Z5 and M5 in \citet{bai11a})
and with STS turned on. In these simulations $N_{\rm STS}$ reaches 12, and
the stress level we find is the same as reported in \citet{bai11a}.
Combining our tests with the successful tests of \cite{osullivan06}, \cite{osullivan07}, and
\cite{choi09}, we are confident that the STS technique implemented here is
capable of achieving substantial speed-up while maintaining accuracy.

\end{document}